\documentclass[aps,pre, twocolumn]{revtex4-1}
\usepackage{graphicx}
\usepackage{amsmath}
\usepackage{amsfonts}
\usepackage{amsthm}
\usepackage{amssymb}
\usepackage{amsbsy}
\usepackage{wasysym}
\usepackage{bm}
\usepackage{mathrsfs}
\usepackage{color}
\usepackage{times}
\usepackage[resetlabels]{multibib}
\usepackage{mathtools}
\usepackage{subfigure}
\usepackage{caption}
\usepackage{dcolumn}
\usepackage[utf8]{inputenc}
\usepackage[english]{babel}
\usepackage{verbatim}
\begin{document}

\title{Mechanical Properties and Pore Size Distribution in Athermal Shear-Strained Porous Glasses}
\author {Sucharita Niyogi and Bhaskar Sen Gupta}
\affiliation{Department of Physics, School of Advanced Sciences, Vellore Institute of Technology, Vellore, Tamil Nadu - 632014, India}

\date{\today}
\begin{abstract} 
 In this paper we study the mechanical properties and pore structure in a three-dimensional molecular dynamics model of porous glass under athermal quasistatic shear. The vitreous samples are prepared by rapid thermal quench from a high temperature molten state. The pore structures form via solid-gas phase separation. The quiescent samples exhibit a wide range of pore topography, from inter-connected pore network to randomly distributed compact pores depending on the material density. We find the shear modulus strongly depends on the density and porosity. Under mechanical loading, the pore structure rearranges which is reflected in the pore size distribution function. Our results show that with increase in strain the distribution widens as the adjacent pores coalesce and form larger pores. We also propose a universal scaling law for the pore size distribution function which offers excellent data collapse for highly porous materials in the undeformed case. From the data scaling we identify a critical density which can be attributed to the transition point from a porous-type to bulk-type material. The validity of the scaling law under finite deformation is also analyzed. 
\end{abstract}
\maketitle
\section{Introduction} 
Porous glass is a specific type of glassy material that includes pores usually in the size range between nanometers to micrometers. These materials have attracted substantial interest in the field of research and industry due to their diverse applications starting from biomedical implants: such as tissue engineering and drug delivery, energy storage and conversion, functional applications in the process of heat conduction, civil infrastructure including wear resistance tools  \cite{yang,porter,Li,zhang,hammel,fu}. The physical properties of porous glass largely depend on the pore structure and distribution. Therefore, it is imperative to understand the relationship between the mechanical properties of such materials with the pore structure to prepare smarter materials.\\
In recent years extensive experimental and numerical studies have been carried out to understand the relationships between the properties of porous glasses with their microstructure. For example it was shown both experimentally and numerically that in metallic porous glass, under tensile loading the shear band formation takes place via the strain localization along the direction of periodic arrangement of pore \cite{sarac,sarac1,sopu}. On the contrary, when tensile loading is applied, the material hardens via the destruction of pores, resulting the inhibition of shear band propagation \cite{wang1}. Using the large-scale MD simulation on nanophase silica glasses it was observed that the pores are having a self-similar structure with a fractal dimension close to 2 \cite{campbell}. Furthermore, the short-range structures of these materials are similar to the bulk glasses whereas in the intermediate-range order it is very different from the bulk. The elastic modulus was found to vary with density as a power law with exponent 3.5 \cite{campbell}. Later MD simulations revealed that nanoporous silica with 50\% porosity, the elastic moduli–porosity relationship can be fitted with either power-law or exponential function \cite{rimsza}. Very similar to ductile metallic alloys, at high strain in porous silica glasses nanocracks were observed to appear on void surfaces resulting in the ligaments fracture because of the growth and coalescence of ligament nanocavities \cite{chen}. \\
Recent MD simulation on microtopography of binodal glasses showed that the pore size distribution strongly depends on porosity \cite{testard,testard1,maxim}. Furthermore, substantial applications of porous materials show the core involvement of guest molecules via absorption method inside the pores \cite{ongari}. Because of this reason, void space in porous glasses plays a crucial part in material characterization. The athermal quasistatic simulation was employed to study the structural evolution of porous glasses at very low temperature under mechanical straining \cite{michele}. A significant change in the structure of porous was observed via the creation of large-scale voids and resulting in a tougher material formation. But despite these extensive research, we are far from a detailed understanding of the mechanical properties of these  porous structures and their relationship with the elastic, shear and bulk moduli.\\
In this paper we study numerically the pore size distribution and mechanical properties of a model porous glass under simple shear deformation. The porous glass is prepared via a sudden thermal quench from the equilibrated liquid state to a very low temperature ($T\rightarrow0$) using MD simulation. The mechanical deformation is applied via the athermal quasistatic process where the system is free from any thermal fluctuation. A wide range of pore morphology is observed which is dependent on the average density of the material. A significant change in the pore size distribution is observed with deformation. In some cases small pores coalesce and form large pores of the order of system size. The shear modulus and porosity show a power-law dependence on the density.  We find the pore size distribution shows a scaling law for the undeformed system. Our result indicates that at high density the system with randomly scattered isolated pores can be characterized by the Gaussian distribution. In the sheared samples we find the distribution function gradually deviates from the scaling law at the intermediate and small length scale with increase in strain.

The paper is organized as follows. In the next section we outline the model and the numerical method to prepare the porous glass and the deformation protocol. In section III we present results obtained from our simulation. We discuss about the formation of porous glass with various pore topography. The relation of the shear modulus with the system density and porosity is investigated. The characterization of the porous structure is demonstrated in terms of pore size distribution function. The scaling law of these distributions is examined under various loading conditions. Finally, in section IV, we offer a summary and a discussion of the results presented in this paper.
\section{MODEL AND NUMERICAL METHODS}
\subsection{Molecular Dynamics Simulation}
In this present work, we carry out MD simulation to prepare the porous glass using the well studied Kob Andersen (KA) model \cite{kob} of a $80:20$ binary Lennard-Jones (LJ) mixture in three dimensions in the NVT ensemble. We label $80\%$ of the particles as type A and the rest as type B. Our simulation consists of total 300000 particles to avoid any system size effect \cite{testard}. The interaction potential for a pair of particles has the following form,
\begin{eqnarray}
U_{\alpha\beta}(r) &=& 4\epsilon_{\alpha\beta}\Big[\Big(\frac{\sigma_{\alpha\beta}}{r}\Big)^{12} - \Big(\frac{\sigma_{\alpha\beta}}{r}\Big)^{6} + A_0 \nonumber\\ &+& A_1\Big(\frac{r}{\sigma_{\alpha\beta}}\Big) + A_2\Big(\frac{r}{\sigma_{\alpha\beta}}\Big)^2\Big] \ , r\le r_{cut} \nonumber \\
&=& 0, \,~~~~~~~~~~~~~~~~~~~~~~~~~~~~~~~~ r\textgreater r_{cut} \label{Uij}
\end{eqnarray}
where $\alpha, \beta \in  \rm{A, B}$. The inter-atomic potential parameters are chosen as follows:  $\sigma_{AA} = 1.0, \sigma_{BB} = 0.88, \sigma_{AB} = 0.8$ and $\epsilon_{AA} = 1.0, \epsilon_{BB} = 0.5, \epsilon_{AB} = 1.5$. These parameters ensure stable glass formation and avoid crystallization. The cutoff radius for the LJ potential is set as $r_{cut} = 2.5$ to enhance computational efficiency. For simplicity, the mass of both types of particles m is taken to be the same and equal to 1.  The units of various quantities in our simulation are as follows: lengths are expressed in the unit of $\sigma_{AA}$, energies in the unit of $\epsilon_{AA}$, time in the unit of $\tau=(m\sigma_{AA}^2/48\epsilon_{AA})^{1/2}$ and temperature in the unit of $\epsilon_{AA}/k_{\rm B}$. Here, $k_{\rm B}$ is the Boltzmann constant which is unity. During MD simulation the position and velocity of particles are updated using the velocity-Verlet integration technique \cite{verlet}. The temperature of the system is controlled using Nose-Hoover thermostat \cite{nose}. Also, periodic boundary condition is applied in all directions.

In order to prepare the porous glass samples we first equilibrate the mixture at high temperature $T=1.5$. Subsequently, the system is quenched to the final temperature $T=0.001$ which is sufficiently low to eliminate any appreciable thermal effects and allowed to evolve for a time interval of $2\times10^3\tau$. As a result of sudden quench, phase separation and solidification of the system happens and the pore structure forms across the glassy sample. Finally, the system is brought to the minimum energy state using conjugate gradient energy minimization algorithm where the temperature is formally $T=0$.
\subsection{Athermal Quasistatic Simulation}
To investigate the porous glass under simple shear deformation, we use athermal quasistatic simulation (AQS) with the limit $T\rightarrow0$ and $\dot{\gamma}\rightarrow0$ where $\dot{\gamma}$ is the strain rate.
While numerous computer simulations have been conducted at finite shear rate and finite temperature \cite{gaurav,gaurav1,Liu}, the quasi-static deformation of an amorphous solid at zero temperature has received considerable attention in recent years \cite{Lacks1999_JCP, Barrat2002_PRB, Lemaitre2004_PRL1, Lemaitre2004_PRL2, Lemaitre2006_PRE, Procaccia2009_PRE, Procaccia2010_PRE, Procaccia2011_PRE, santhosh}. The principle utility of the AQS algorithm is that it enables us to probe the shear-induced changes in the geometry of the energy landscape and the system’s trajectory on the same, in the absence of thermal fluctuations. The AQS algorithm includes two iterating steps: (a) the freshly quenched glassy sample is deformed by applying an affine simple shear transformation to each particle $i$ of the system as
\begin{equation}
r_{ix} \rightarrow r_{ix} +r_{iz}\delta\gamma, ~~~r_{iy} \rightarrow r_{iy}, ~~~r_{iz} \rightarrow r_{iz}
\end{equation}
using the Lees-Edwards boundary conditions. We choose sufficiently small strain increment $\delta\gamma=10^{-4}$. (b) After every affine transformation step, the potential energy of the deformed system is minimized using conjugate gradient algorithm under the constraints imposed by the boundary condition. By repeating steps (a) and (b) we can reach up to arbitrarily large strain values. The AQS method ensures that the system is in mechanical equilibrium after every differential strain increase.
LAMMPS simulator package is used to perform all the simulations \cite{lammps}. 
\section{Results}
\subsection{Porous glass formation}
\begin{figure}[htp]
	\renewcommand{\thefigure}{1}
	\centering
	\begin{tabular}{cc}
		\includegraphics[width=42mm]{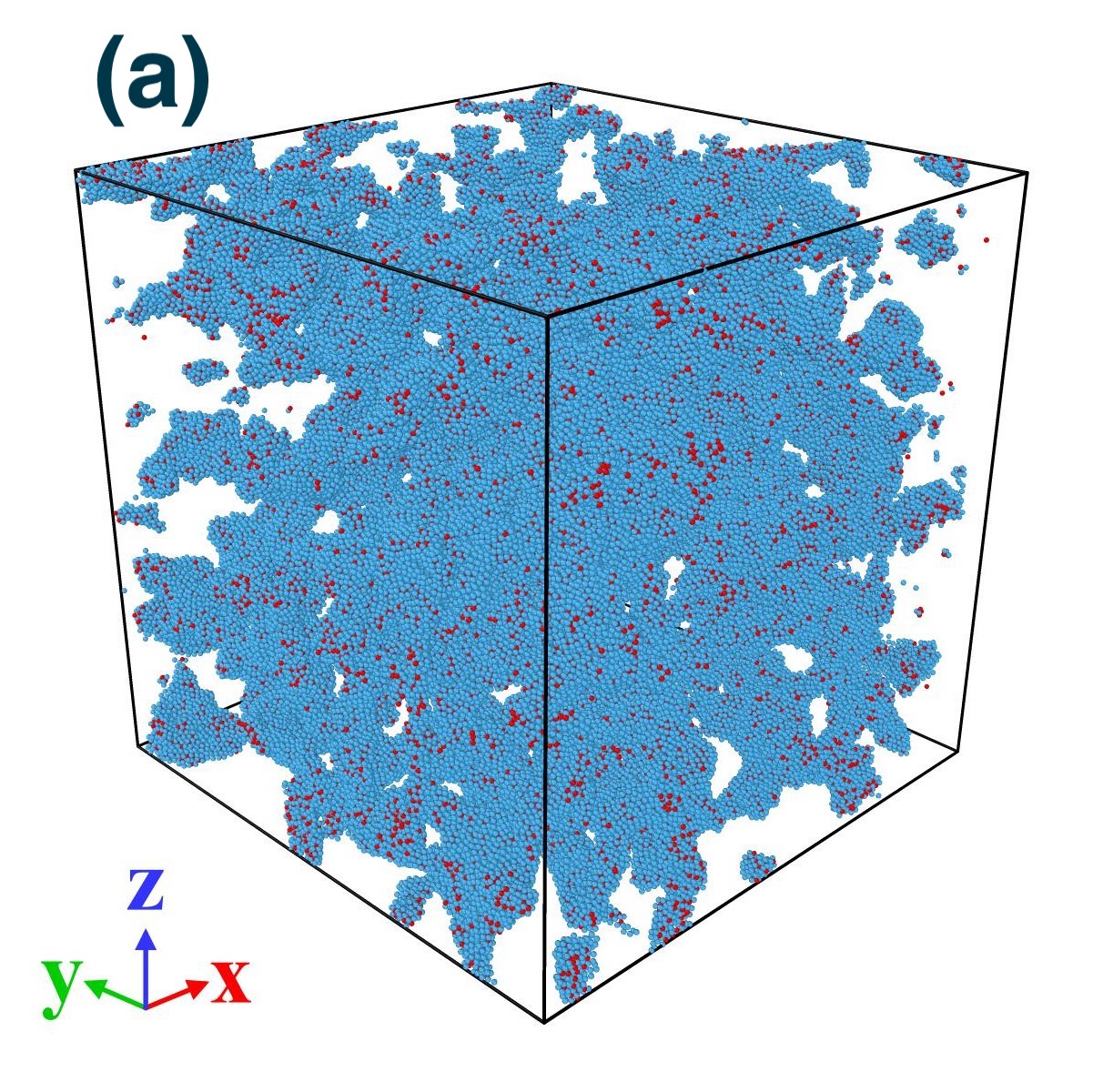}&
		\includegraphics[width=42mm]{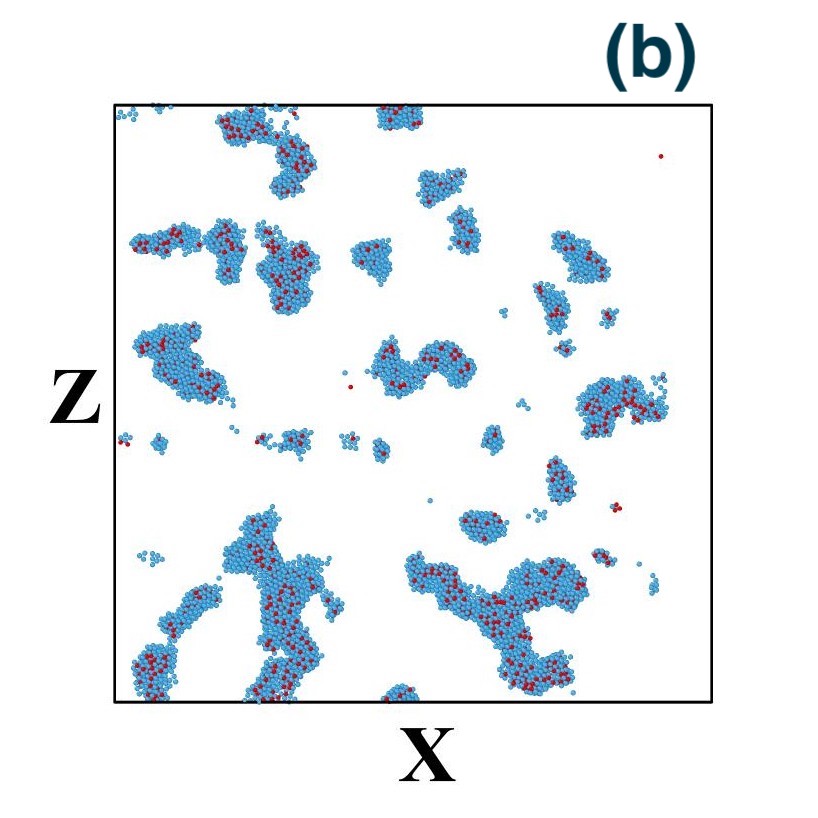}\\
		\includegraphics[width=42mm]{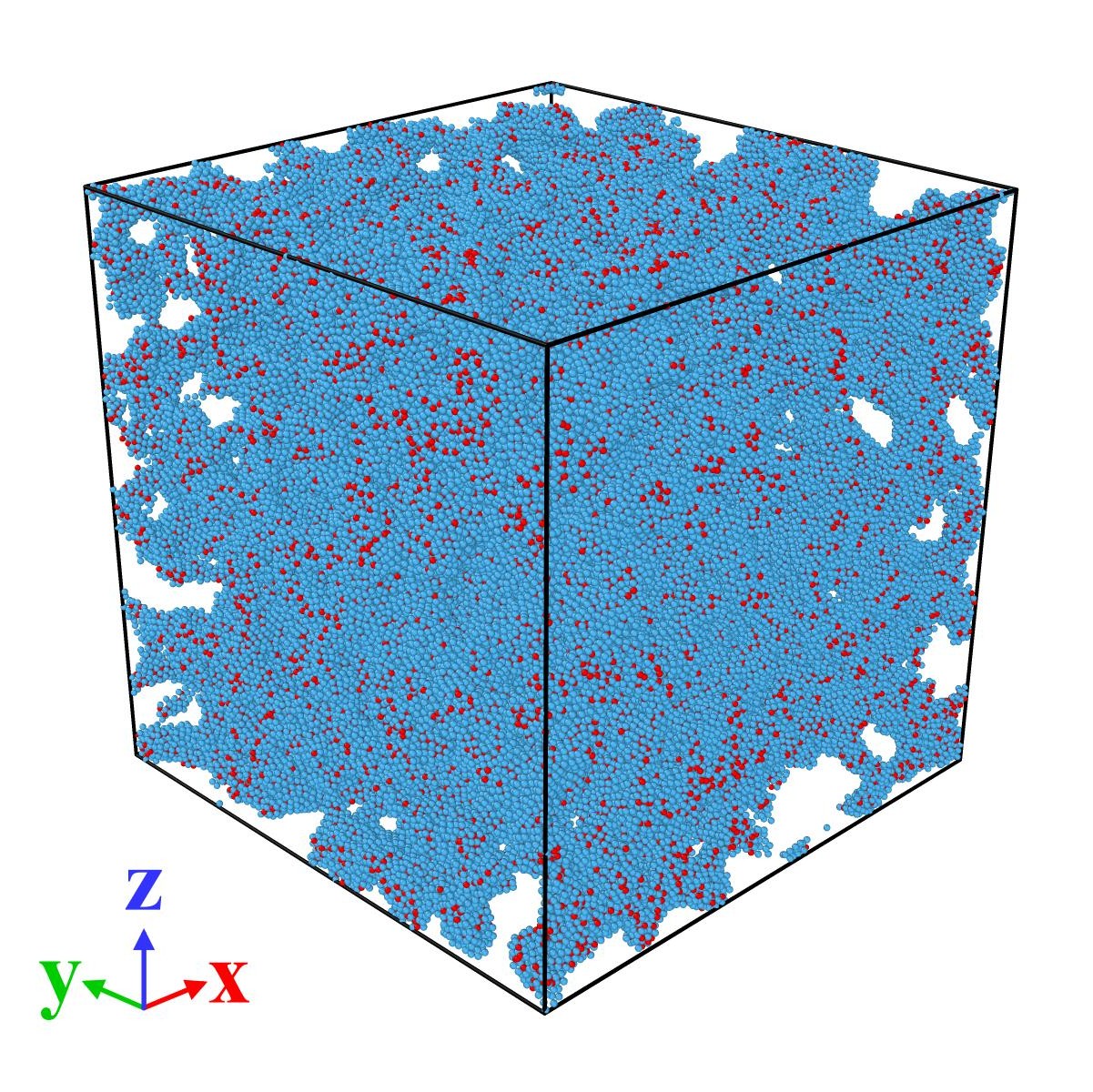}&
		\includegraphics[width=42mm]{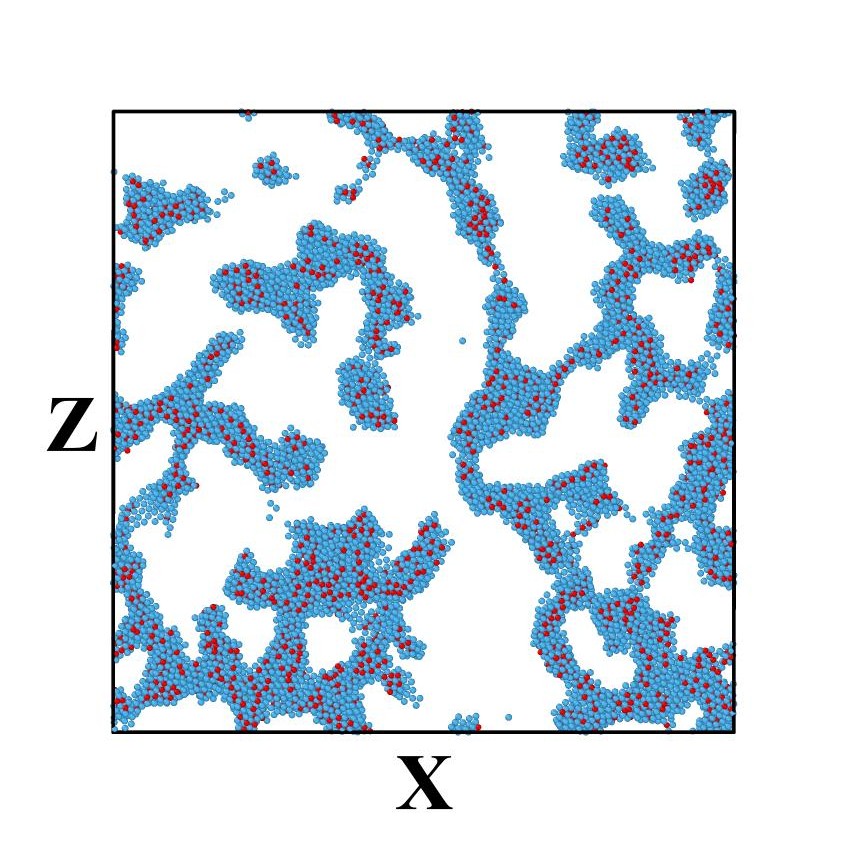}\\
		\includegraphics[width=42mm]{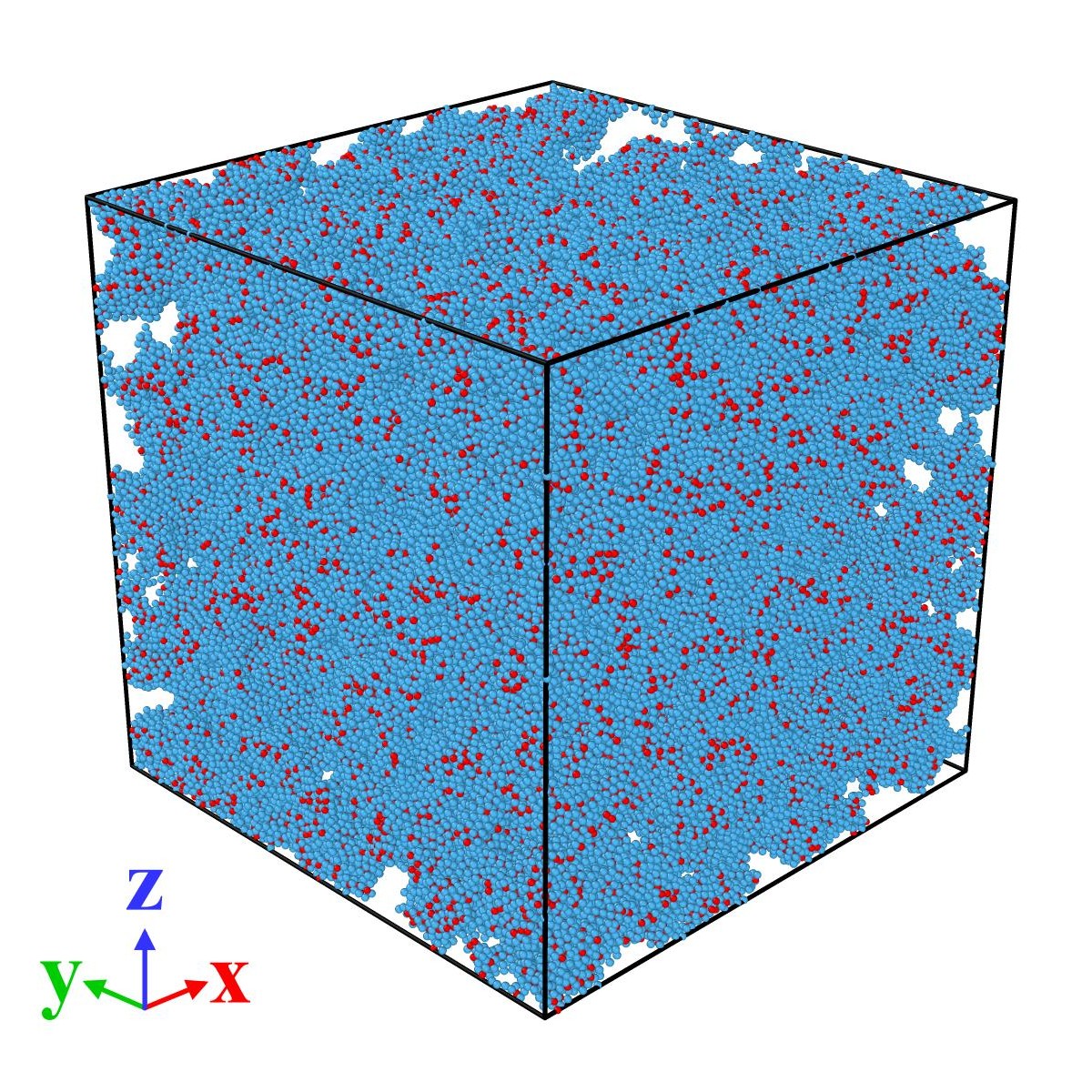}&
		\includegraphics[width=42mm]{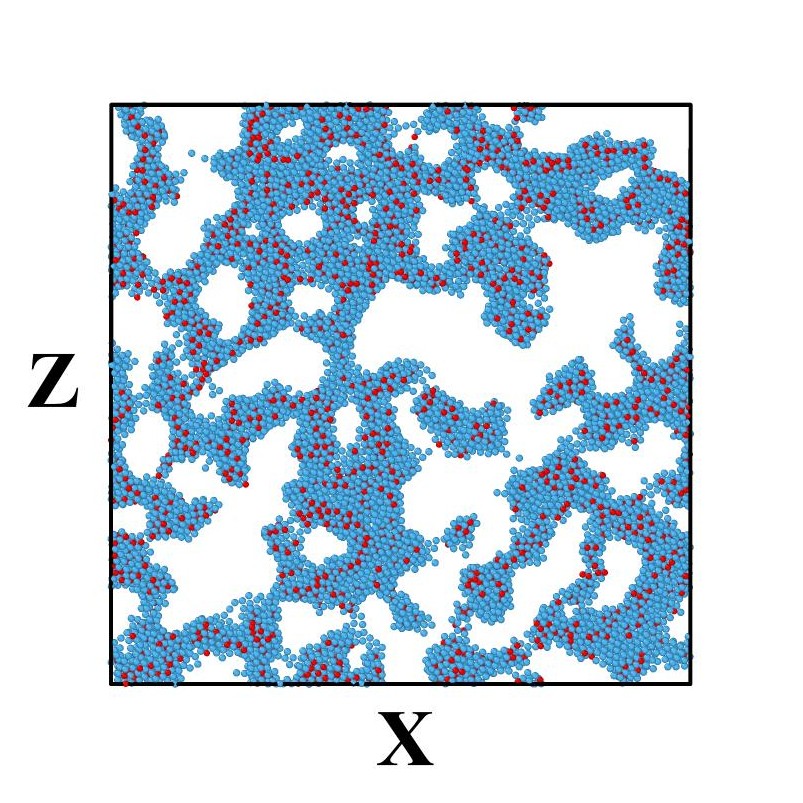}\\
		\includegraphics[width=42mm]{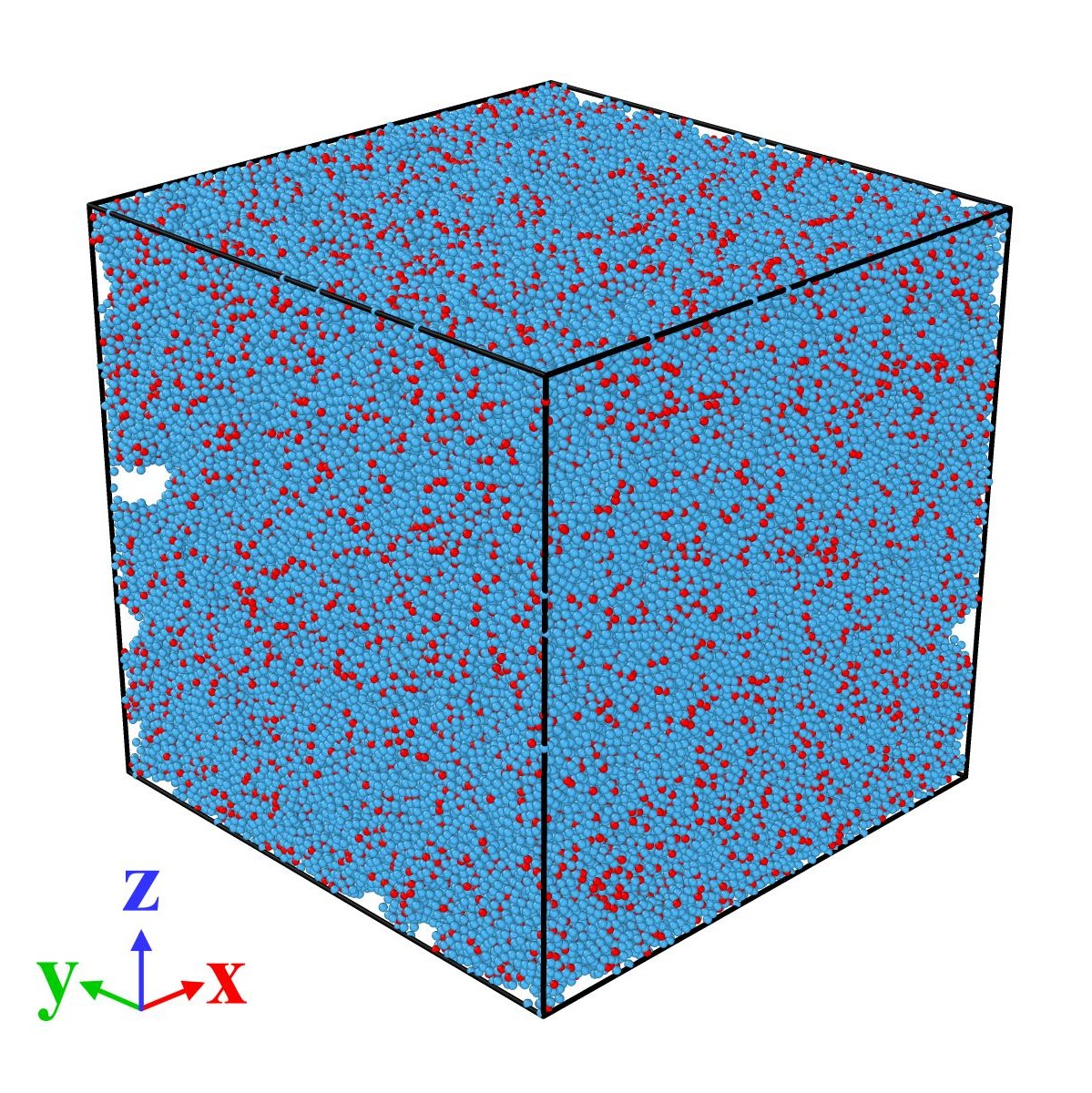}&
		\includegraphics[width=42mm]{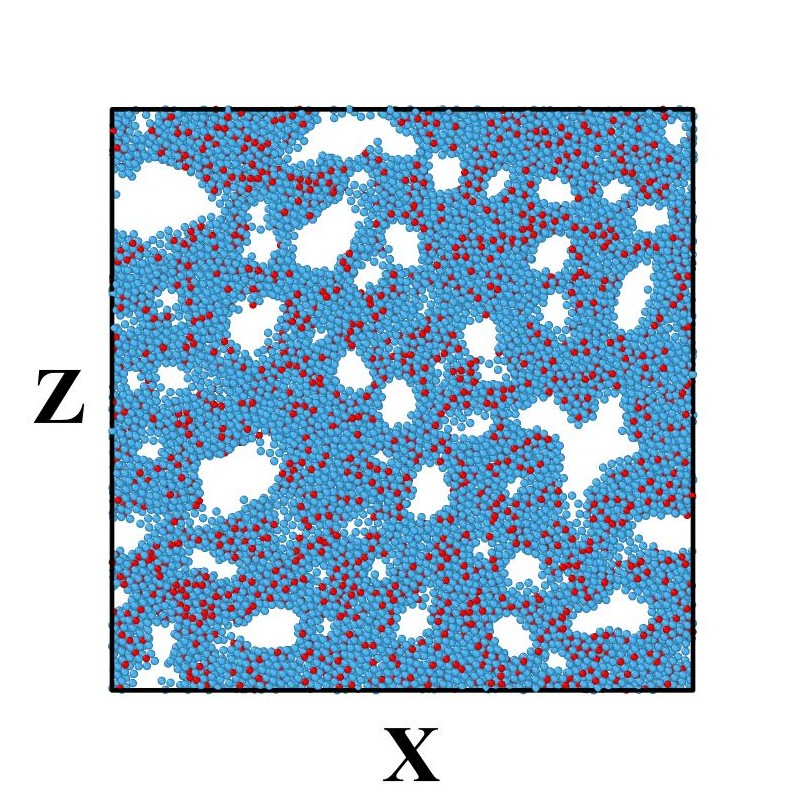}\\
	\end{tabular}
\caption{ (a) Three-dimensional snapshots for the porous glasses at densities $\rho=0.2, 0.4, 0.6$ and $0.9$ (top to bottom). The A and B particles are marked as blue and red respectively. (b) Two-dimensional cross sections of the snapshots in (a).}
\label{config}
\end{figure}

When the glass-forming liquid undergoes a deep thermal quench, the pore structures are developed via solid-gas phase separation procedure. The underlying microscopic mechanisms of phase separation are nucleation and spinodal decomposition. In our study, we equilibrate the system at high temperature and the molten state is quenched suddenly to a very low temperature ($T\rightarrow0$) and is allowed to evolve in time. A fast quench leads to a completely demixed liquid-gas system. During the equilibration at low temperature, the system phase separates into solidified material and pores and eventually porous glass is formed. The material has a bicontinuous morphology and the structural evolution is negligible in the athermal limit. In Fig. \ref{config}(a) we show the typical configurations of the porous glass for a wide range of densities $0.2 \le \rho \le 0.9$ obtained from our simulation. It is challenging to contemplate the three-dimensional complex pore morphology from Fig. \ref{config}(a). This can be better appreciated from Fig. \ref{config}(b), where we present the two-dimensional cross sections of the snapshots in Fig. \ref{config}(a). It is conspicuous that the topographical patterns of the finally evolved pores strongly depend on the average density of the system. At lower density, the pore structures are larger in size and most of the pores are interconnected. A system spanning bicontinuous structure is observed and the volume fraction of the occupied dense phase increases with density up to $\rho=0.8$. For example in Fig \ref{config}(b) we see a complex channel of pores extended over the whole system. With further increase in density, the pore structure entirely changes and we observe randomly distributed isolated individual pores embedded into solid phase. While the experimental investigation is lacking in case of highly porous materials, the findings of random distribution of isolated pores at high density substantiates our simulation results \cite{yavari}. 

\subsection{Deformation of Porous glass}
\begin{figure}[h!]
	\renewcommand{\thefigure}{2}
	\centering
	\includegraphics[width=0.99\columnwidth]{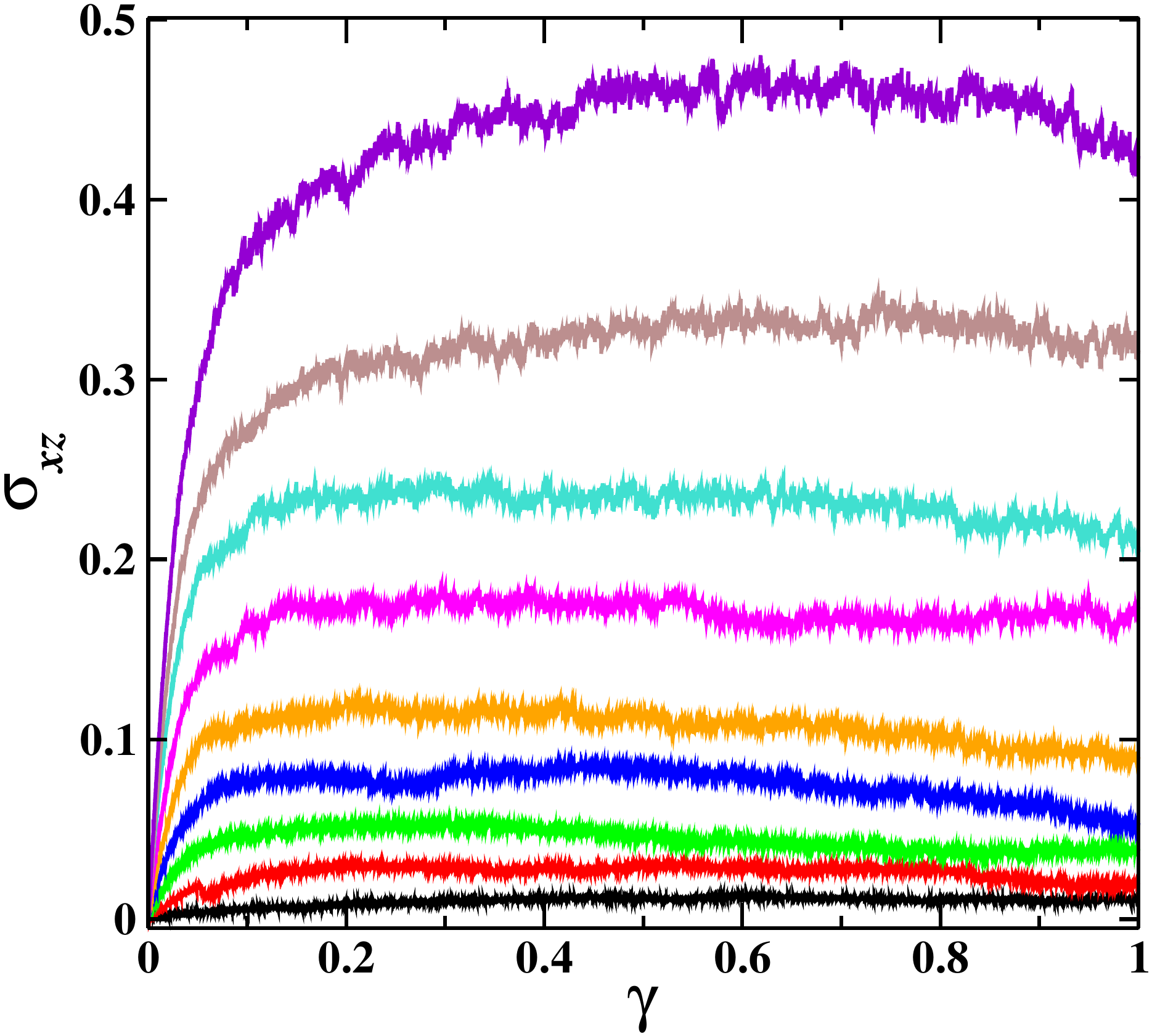}
	\caption{The stress vs. strain curve for our porous glass system under simple shear deformation for the densities $\rho=0.2, 0.3, 0.4, 0.5, 0.6, 0.7, 0.8, 0.9$ and $1.0$ (bottom to top). The results are averaged over $5$ independent runs.}
	\label{stress}
\end{figure}

\begin{figure}[h!]
	\renewcommand{\thefigure}{3}
	\centering
	\includegraphics[angle=-90,width=0.99\columnwidth]{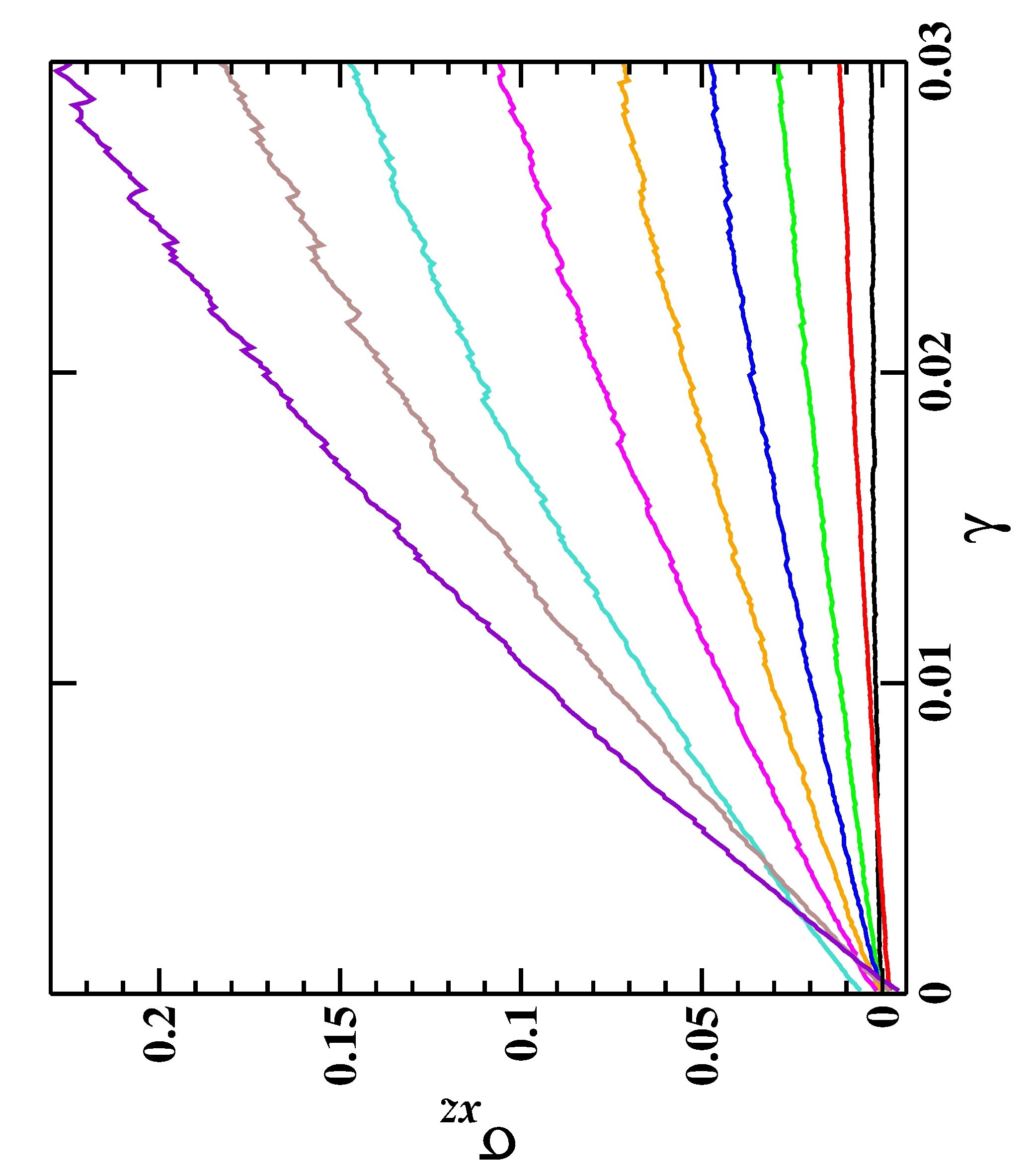}
	\caption{The same stress vs strain curve shown in Fig \ref{stress} zoomed at the small strain region.}
	\label{G}
\end{figure}

\begin{figure}[h!]
	\renewcommand{\thefigure}{5}
	\centering
	\includegraphics[width=0.99\columnwidth]{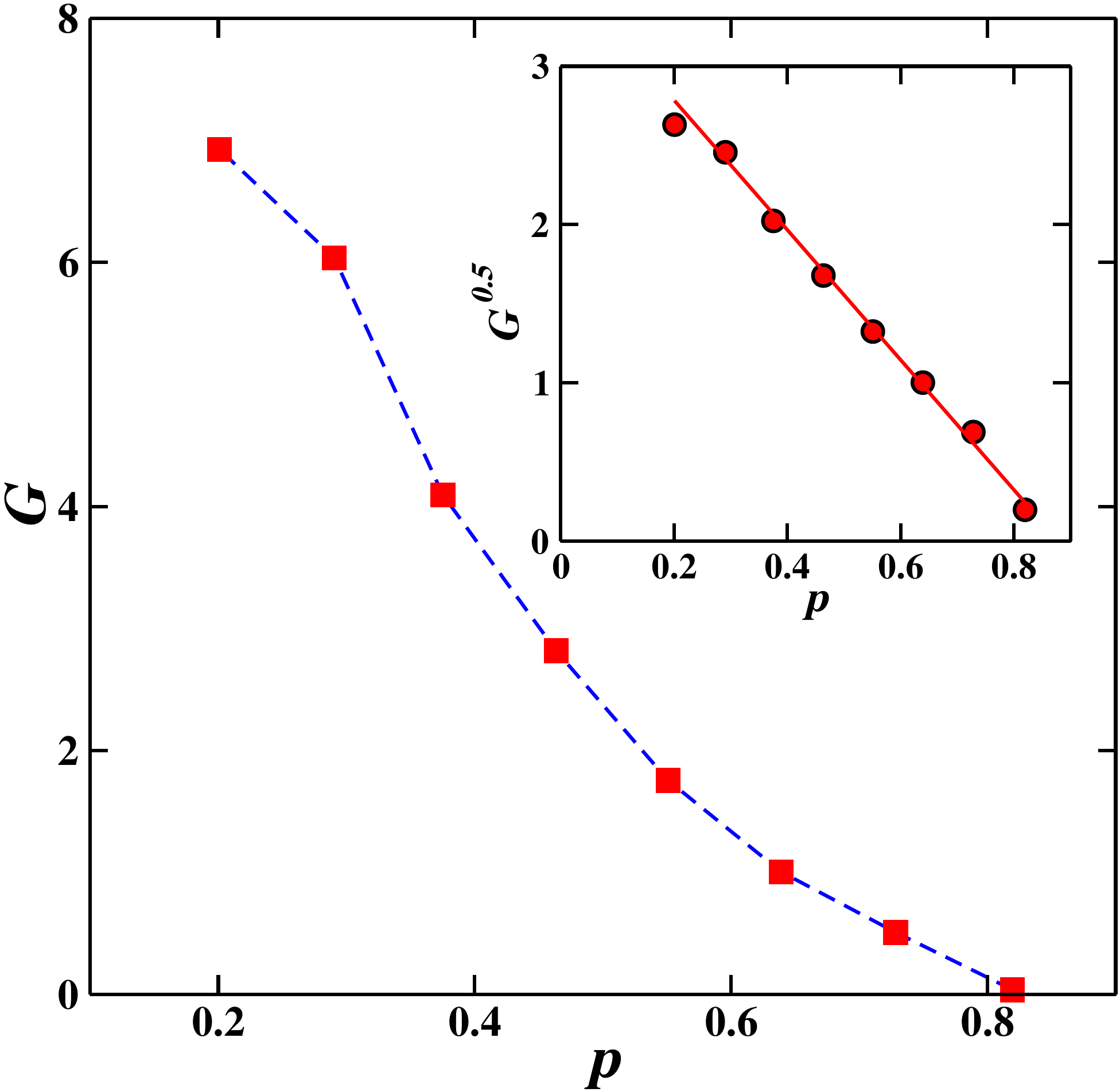}
	\caption{The variation of shear modulus $G$ as a function of porosity. Inset: The same data plotted as $G^{0.5}$ vs porosity with circles. The solid line is the best fit curve.}
	\label{GvsP}
\end{figure}

Once an ensemble of porous glass is prepared at various densities, we strain each sample in the AQS limit ($T\rightarrow0, \dot{\gamma}\rightarrow0$) to examine the stress-strain curve and the shear modulus. We apply simple shear on the $xz$ plane in the direction of $x$. In Fig. \ref{stress} we show the shear stress as a function of shear strain for a set of average glass densities in the range $0.2 \le \rho \le 1.0$. 
The result is averaged over five independent samples. As observed in almost every amorphous solid, the stress here is directly proportional to strain up to a certain deformation, and after that the material yields via shear band formation. On further shearing, the system reaches the steady-state plastic flow. From the figure it is clear that the results strongly depend on the density of the system. Due to the rapid thermal quench we do not observe any stress overshoot near the yielding transition point. \\
Next, we examine the change in the shear modulus of the material with density. For that, we focus on the linear regime of the stress-strain graph at small $\gamma$ value. This is depicted in Fig. \ref{G}. The shear modulus $G$ is computed by measuring the slope of the stress-strain graph in Fig. \ref{G}. It is clear from the figure that the slope increases with density.
\begin{figure}[h!]
	\renewcommand{\thefigure}{4}
	\centering
	\includegraphics[angle=-0,width=0.99\columnwidth]{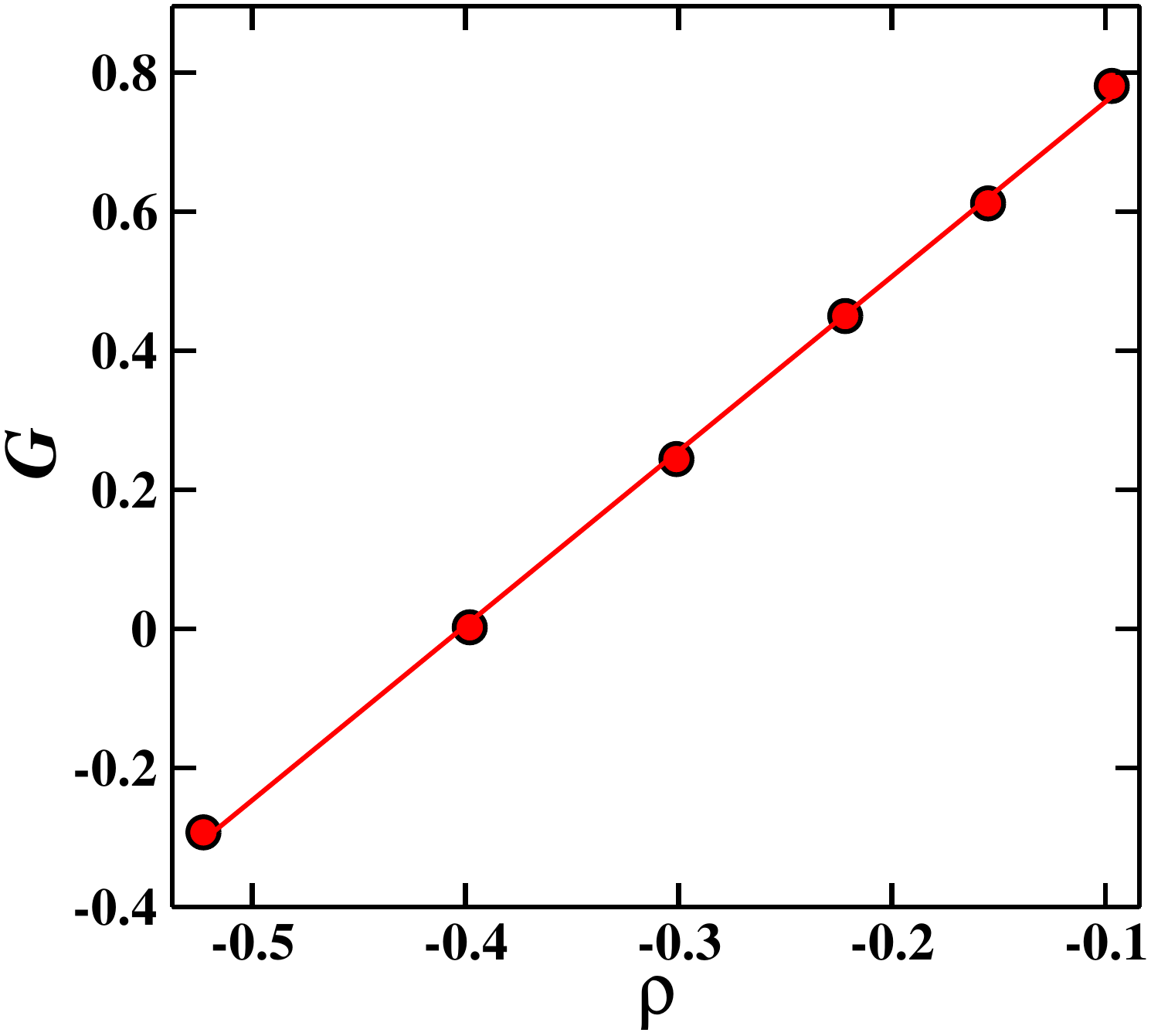}
	\caption{The variation of shear modulus $G$ as a function of density $\sigma$ in the log-log scale is shown by circles. The solid line is the power law fit with exponent $2.5$ (see text).}
	\label{Gfit}
\end{figure}
The $G$ vs density curve is shown in Fig. \ref{Gfit} in the log-log scale. The best fitted curve to this data suggests the scaling law: $G\sim\rho^{2.5}$.\\
For further understanding we examine the variation of shear modulus as a function of porosity. This is shown in Fig. \ref{GvsP}. 
To shed some light on this, we follow the approach of the so-called percolation theory which successfully relates Young's modulus $E$ and the porosity $p$ of the porous material as $E \sim (p_c-p)^f$ \cite{sahimi}. Here $p_c$ is the percolation threshold and $f$ is the critical exponent. A similar relation was also derived by Phani and Niyogi \cite{phani} in a semi-emperical way. According to percolation theory, the shear modulus can also be described by the same equation with critical exponent $f$ being the same for Young’s modulus and shear modulus. For an infinite system in three dimension the theory predicts $f=2.1$. We, therefore, choose the functional form for the shear modulus as $G=G_0(p_0-p)^f$. Fitting this functional form to our simulation data yields $f=2.0$ and $p_0=0.85$. The quality of fitting is shown in the inset of Fig. \ref{GvsP}. Therefore, we observe that the function form of $G(p)$ describes the porosity for $p\gtrsim0.3$. The minute discrepancy in $f$  between the theoretical prediction and our simulation result can be attributed to the following reason. In percolation theory the exponent is estimated for an infinite cluster whereas our result is affected by the finite size of the system. Also the percolation threshold $p_c$ depends on several factors including the system size, preparation protocol, pore structure distribution which in turn influences the value of $f$. Therefore, for real systems, $f$ is considered as characteristic exponent instead of a critical exponent. It is worth mentioning here that the shear modulus obtained from experimental data for some materials is found to differ with the percolation theory prediction \cite{kov,kov1} which requires further investigation.

\begin{figure}[htp]
	\centering
	\begin{tabular}{cc}
		\renewcommand{\thefigure}{6}
		\centering
		\includegraphics[width=48mm]{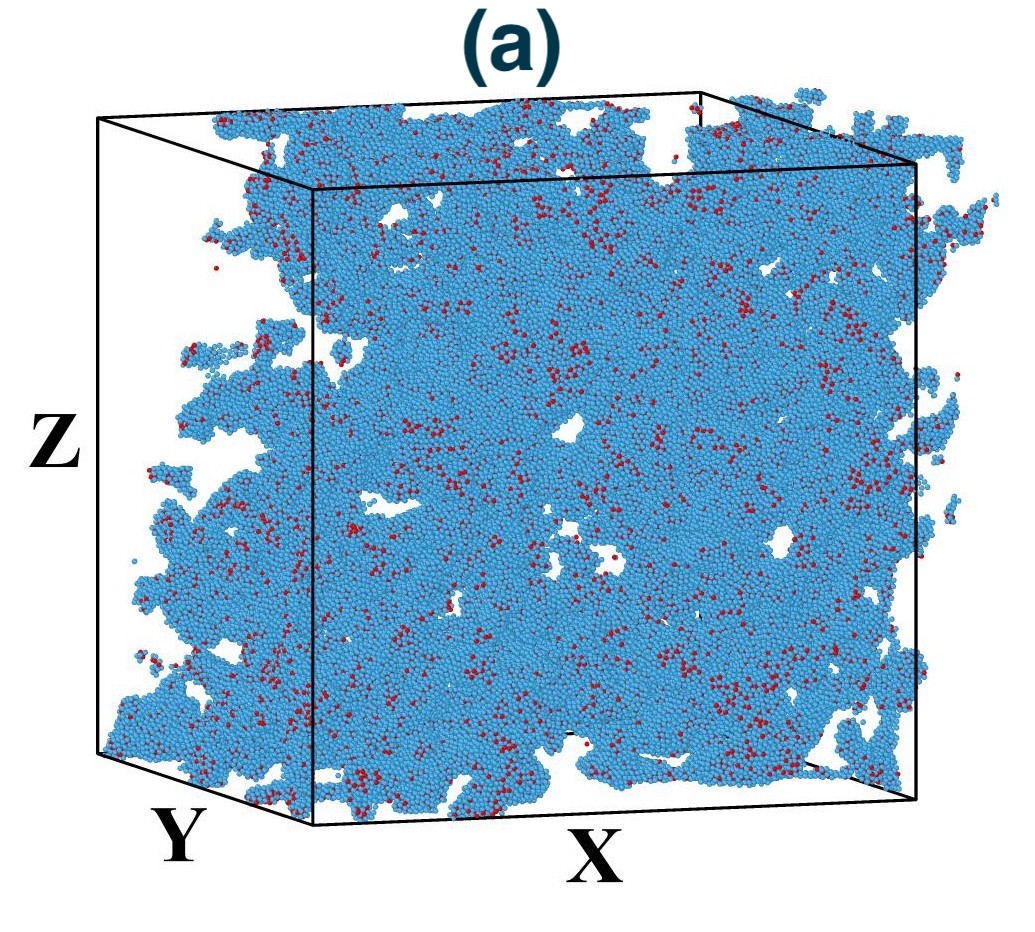}&
		\includegraphics[width=45mm]{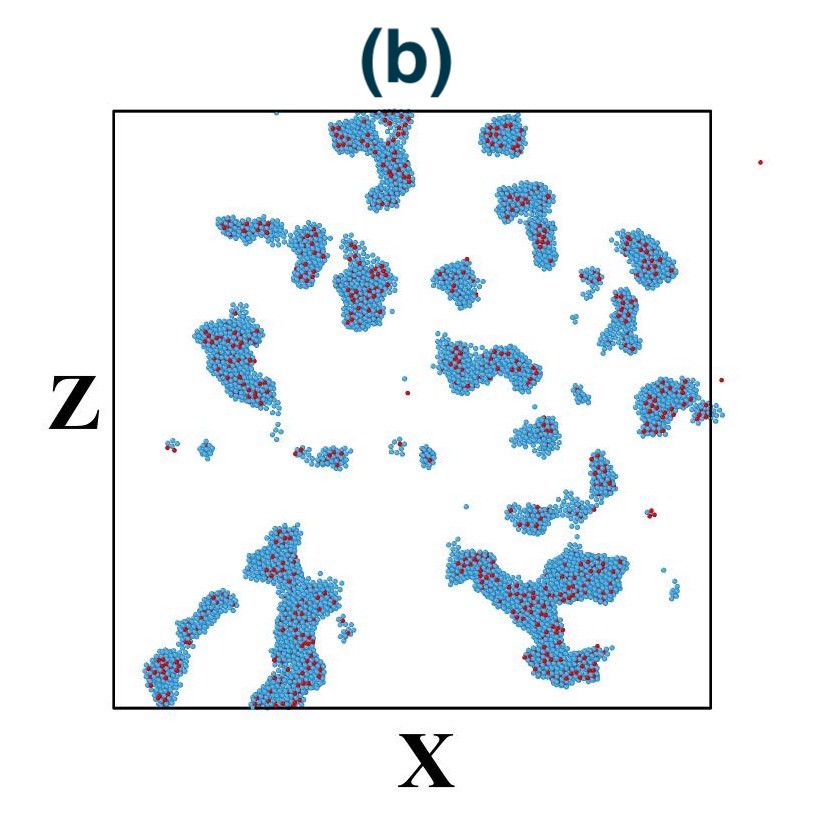}\\
		\includegraphics[width=45mm]{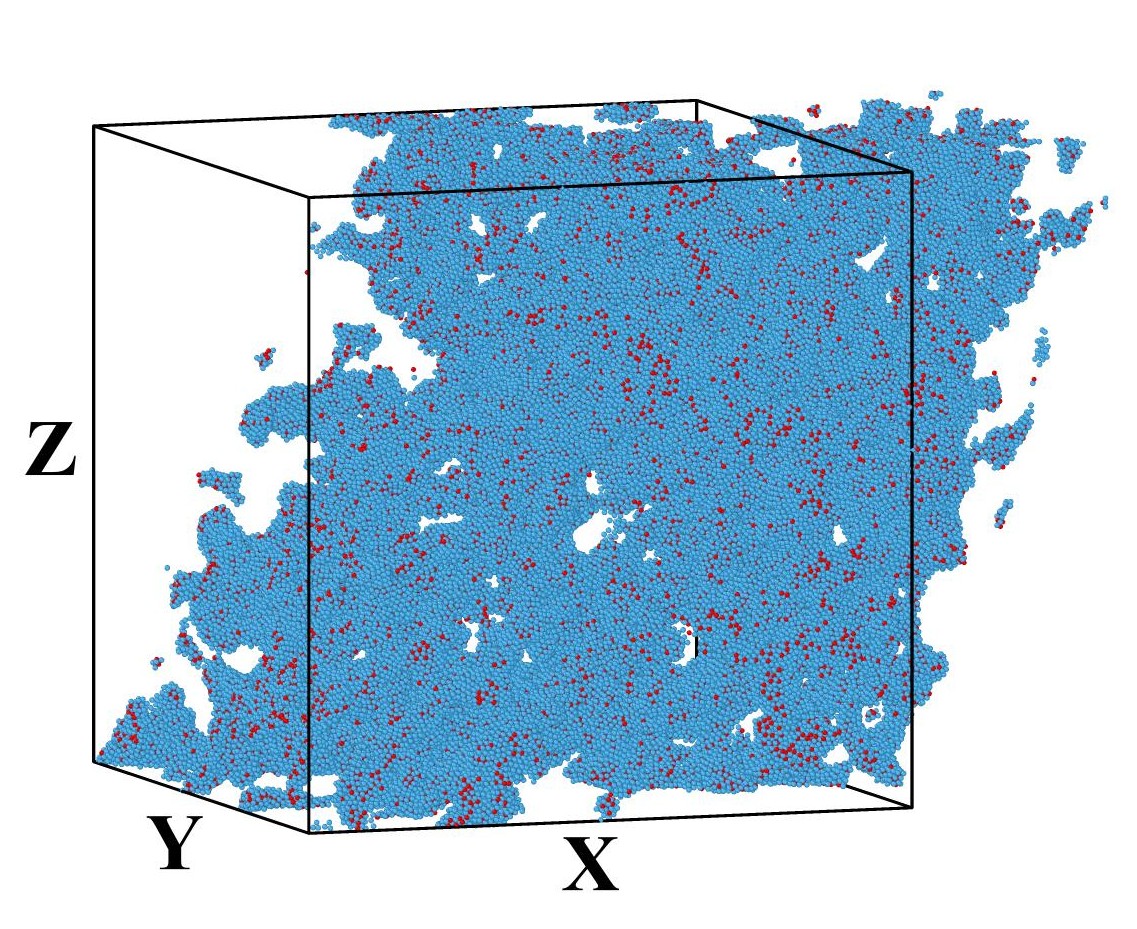}&
		\includegraphics[width=45mm]{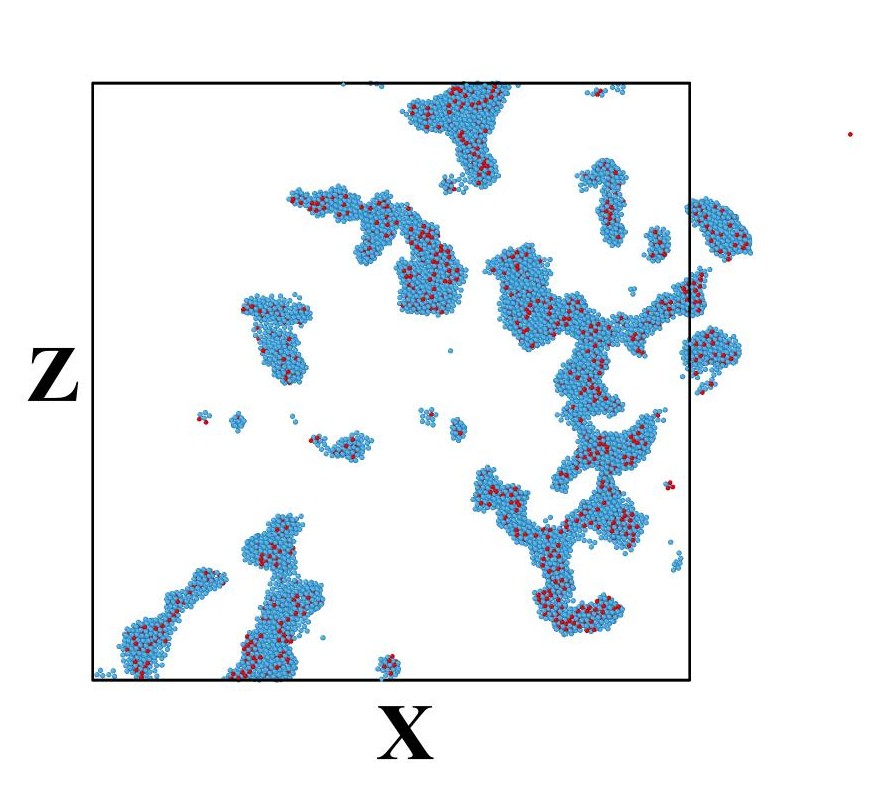}\\
		\includegraphics[width=45mm]{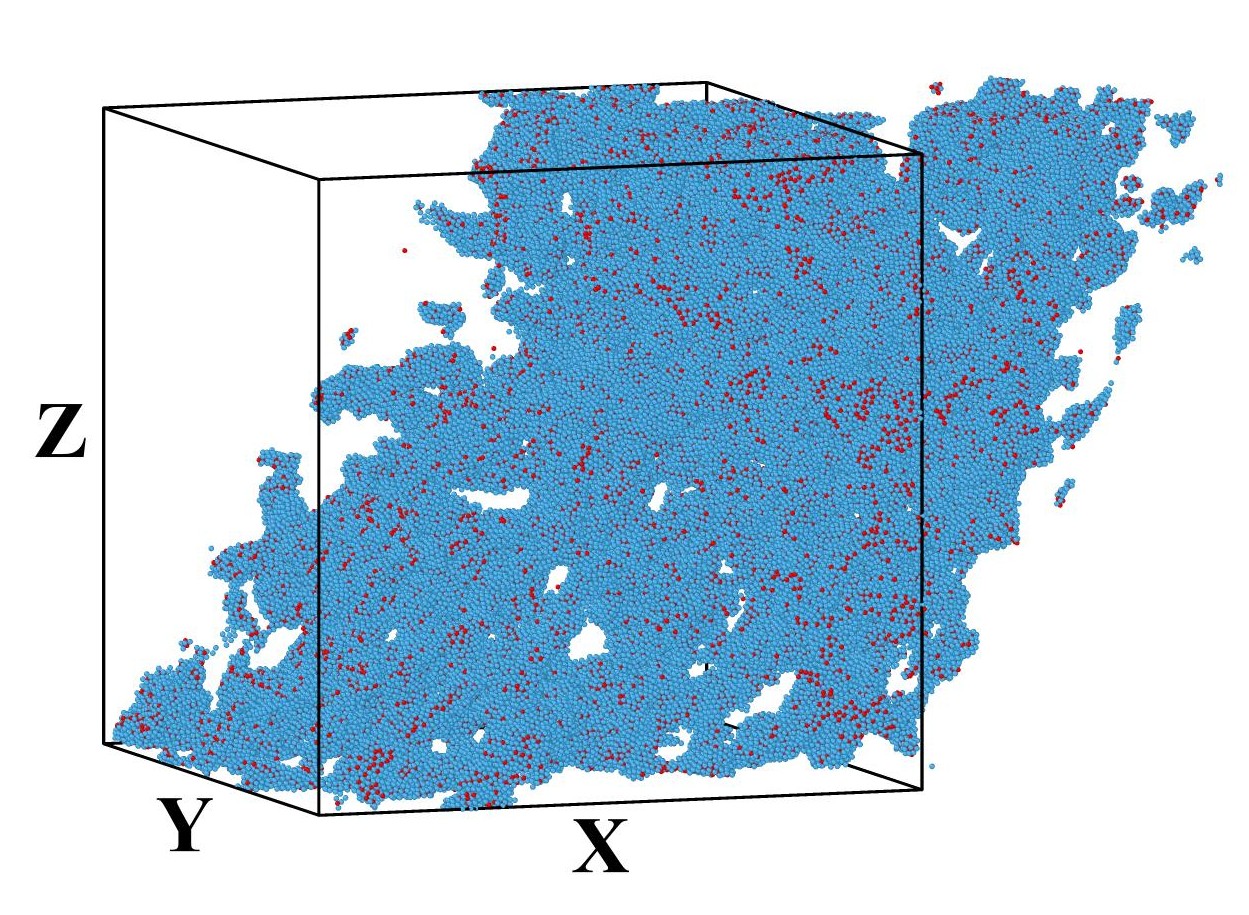}&
		\includegraphics[width=45mm]{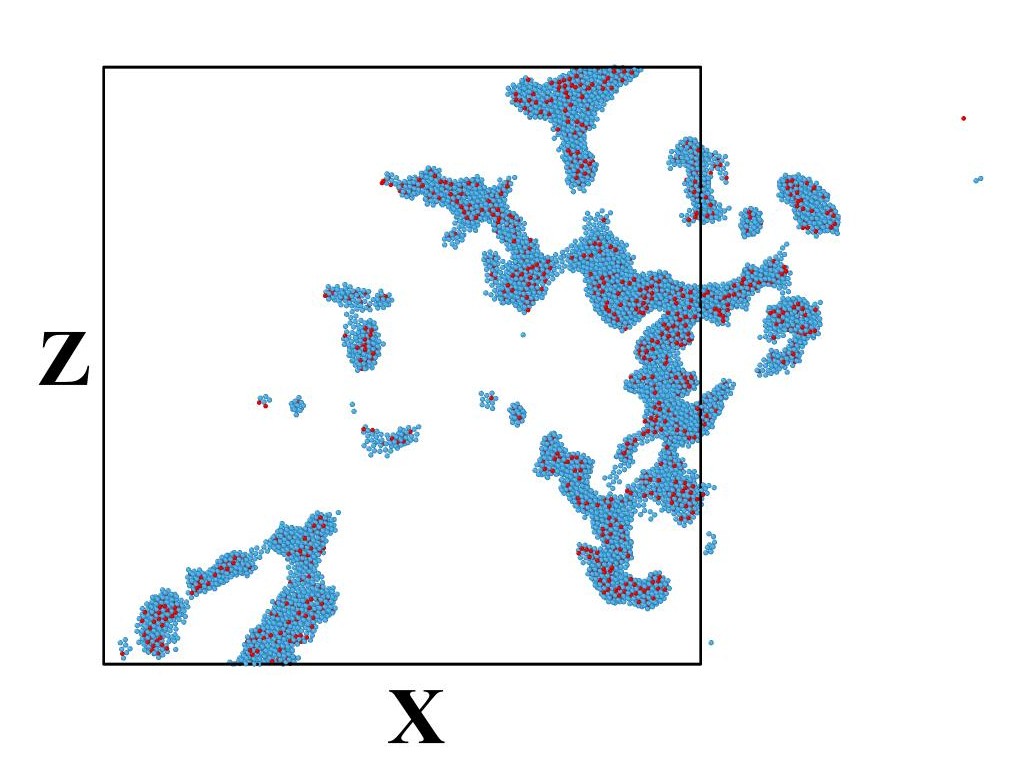}\\
		\includegraphics[width=45mm]{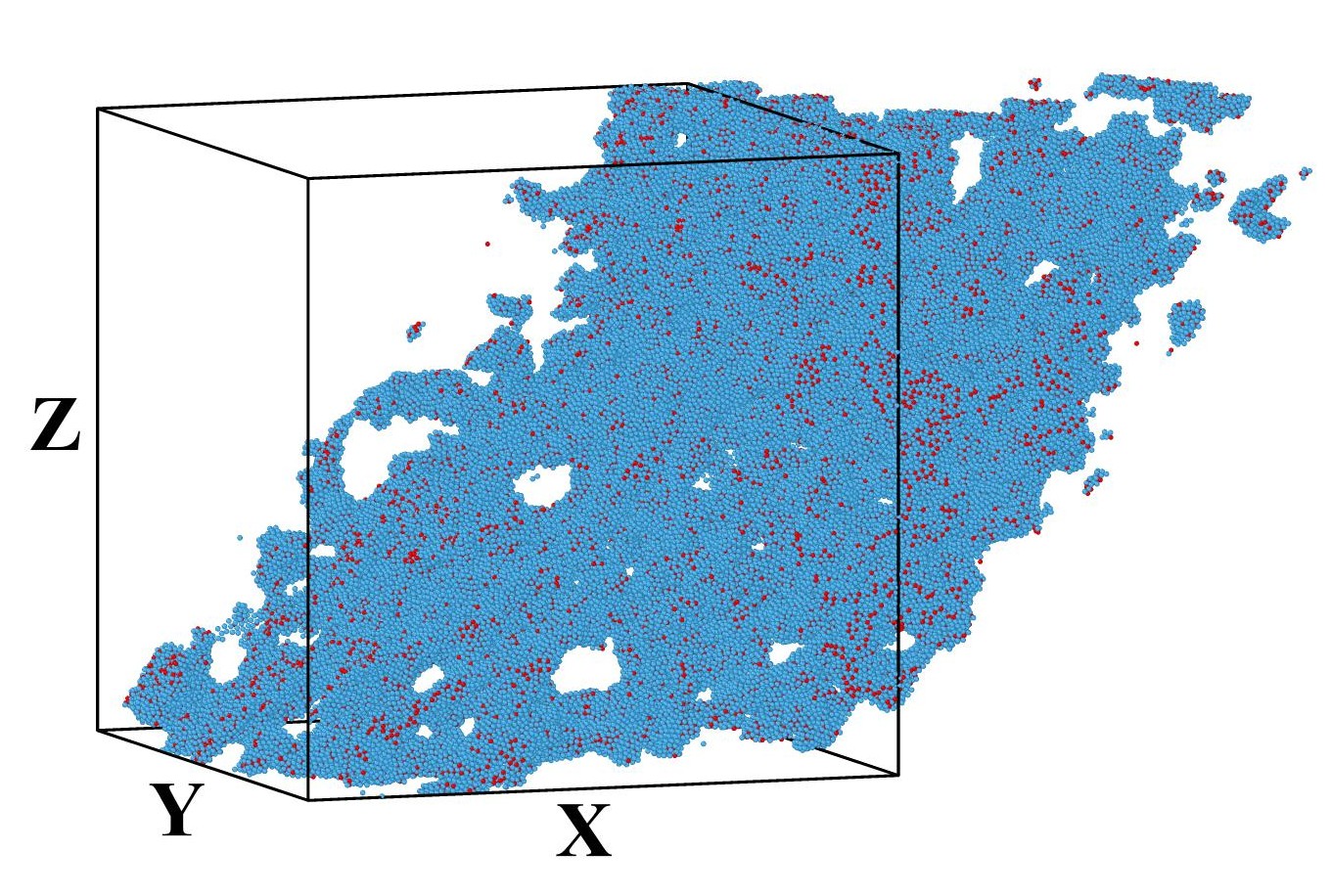}&
		\includegraphics[width=45mm]{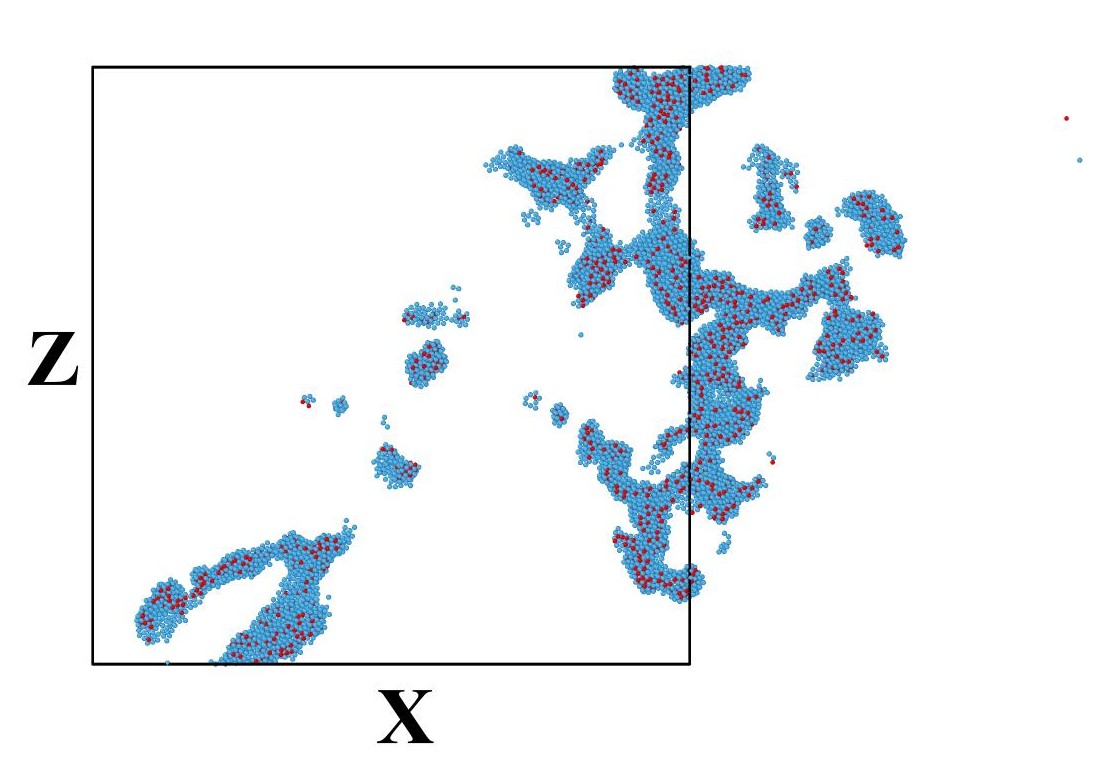}\\
	\end{tabular}
\caption{(a) Three-dimensional snapshots for the porous glasses for the average density $\rho=0.2$ and shear strain $\gamma = 0.2, 0.4, 0.6$ and $0.8$ (from top to bottom). (b) Two-dimensional cross sections of the snapshots in (a).}
	\label{sheared-config-0.2}
\end{figure}

\begin{figure}[htp]
	\centering
	\begin{tabular}{cc}
		\renewcommand{\thefigure}{7}
		\centering
		\includegraphics[width=45mm]{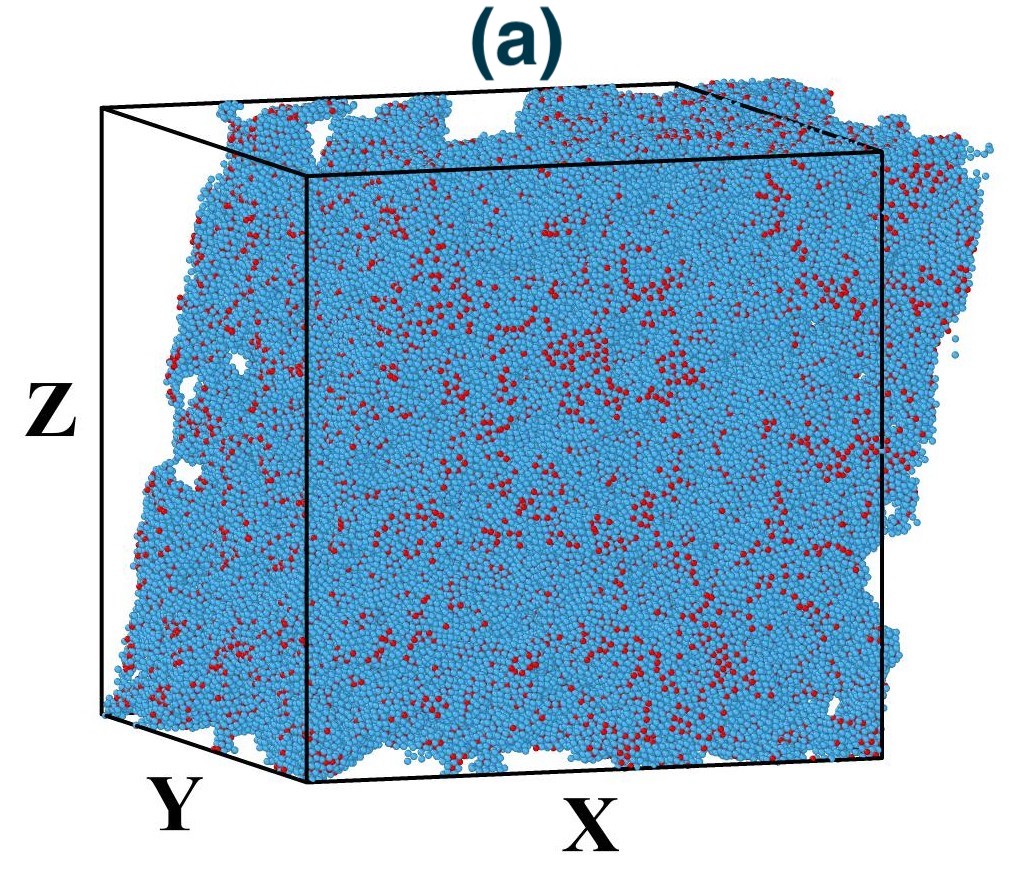}&
		\includegraphics[width=45mm]{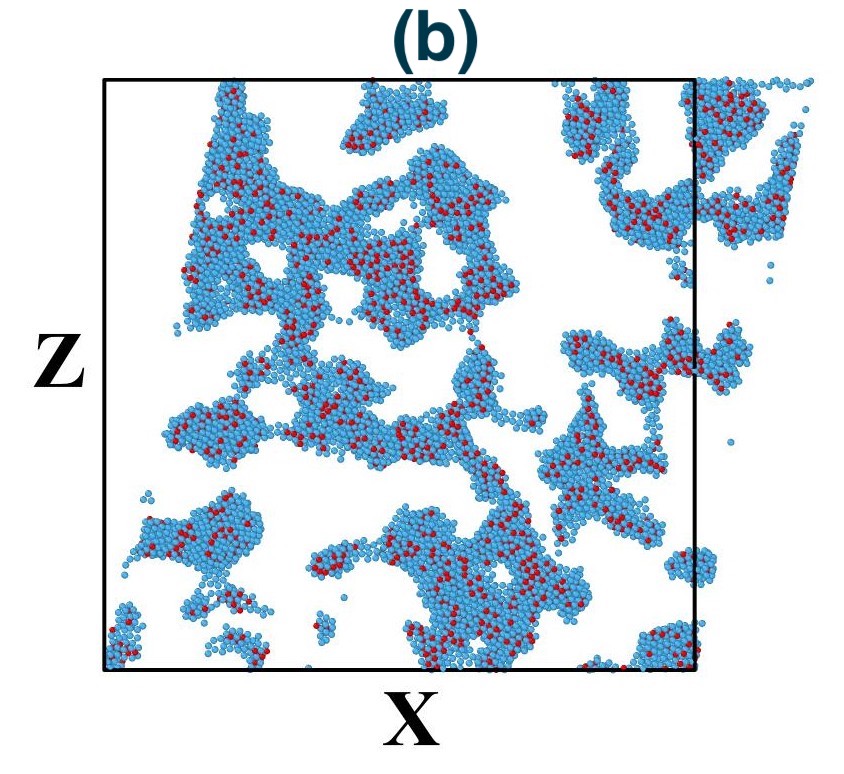}\\\\
		\includegraphics[width=45mm]{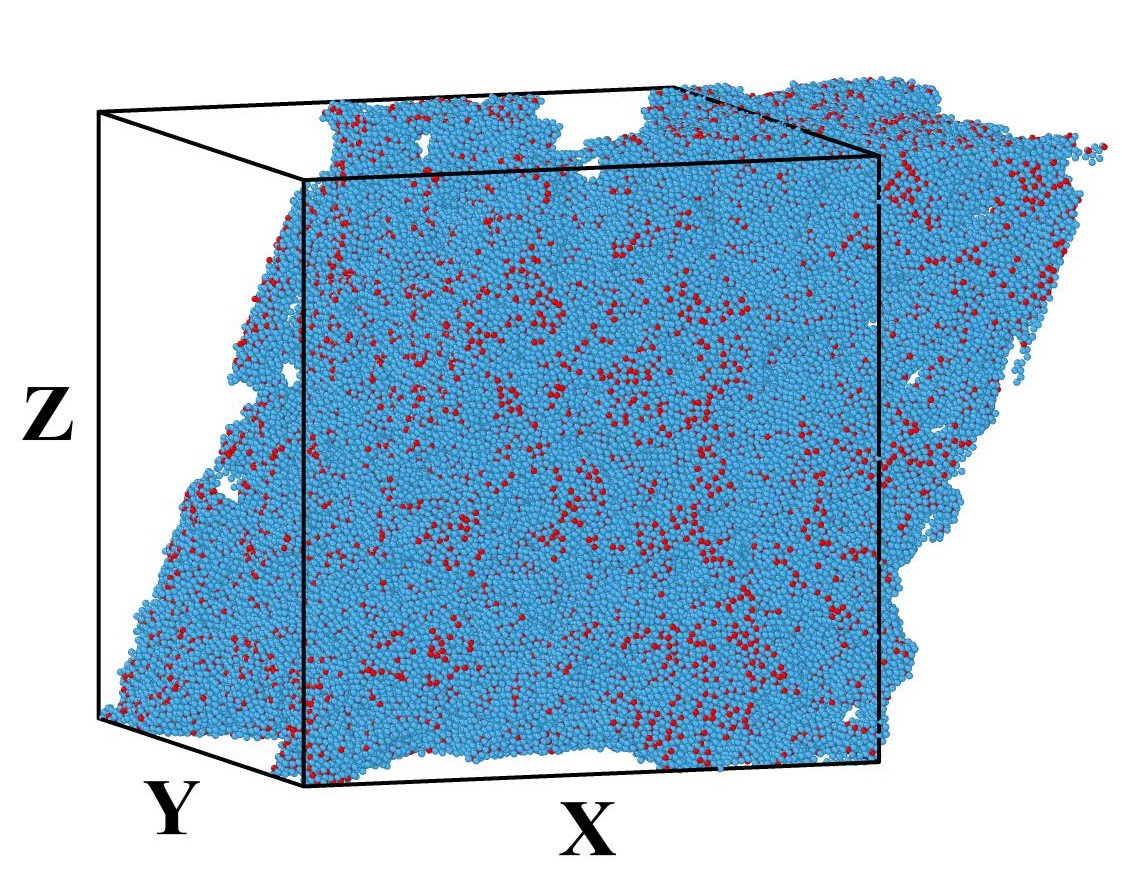}&
		\includegraphics[width=45mm]{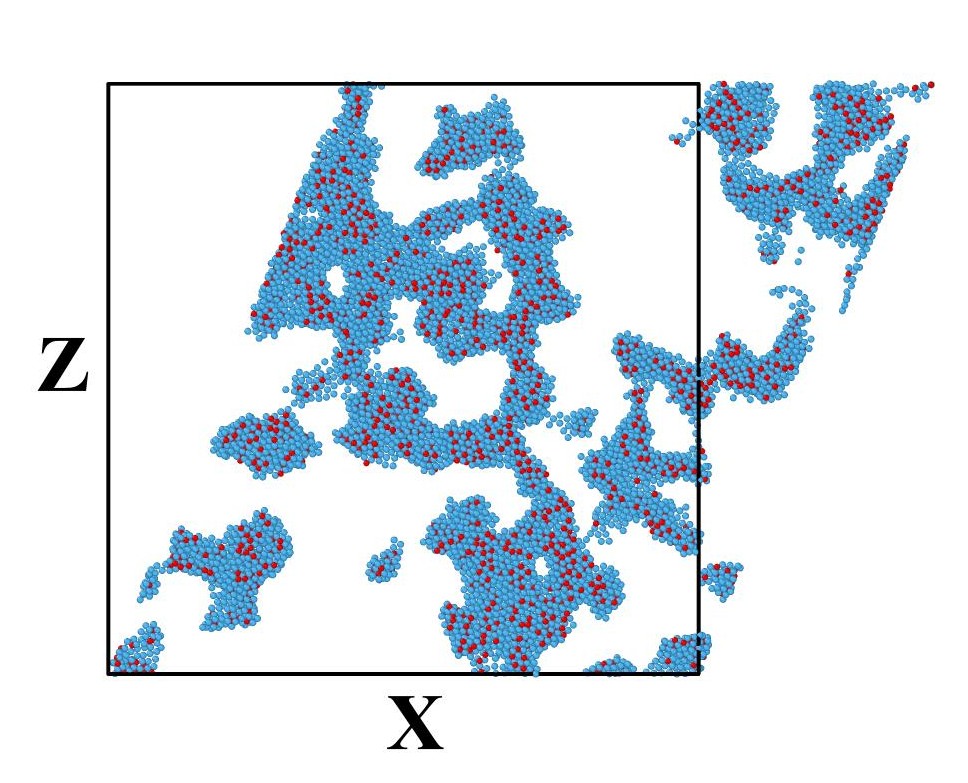}\\\\
		\includegraphics[width=45mm]{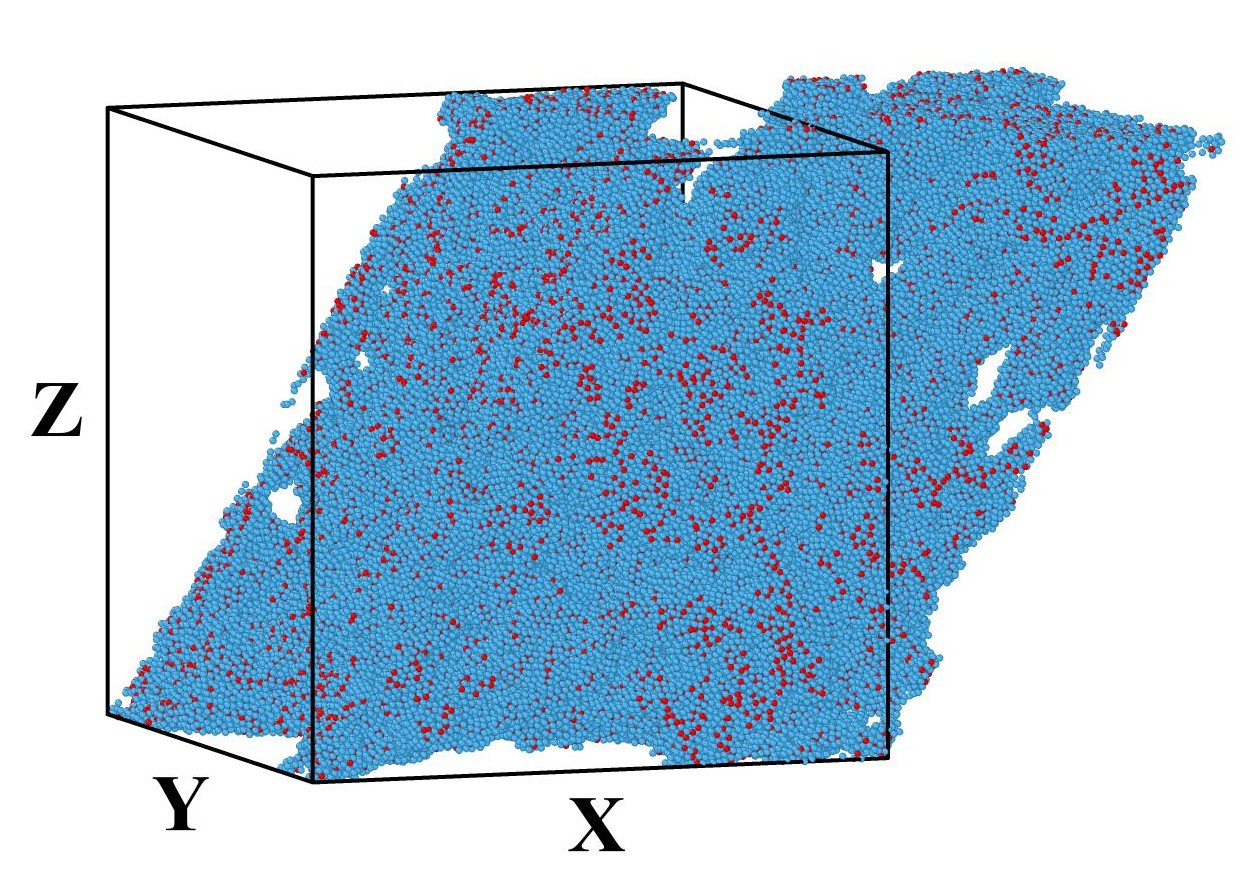}&
		\includegraphics[width=45mm]{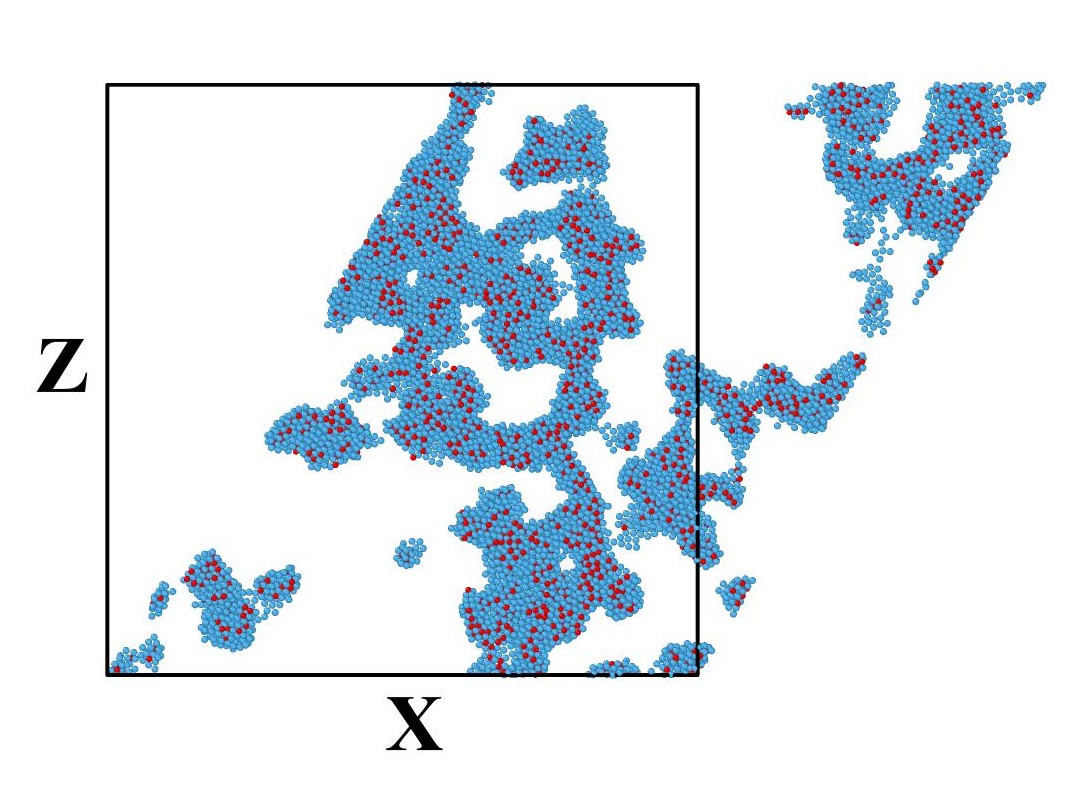}\\\\
		\includegraphics[width=45mm]{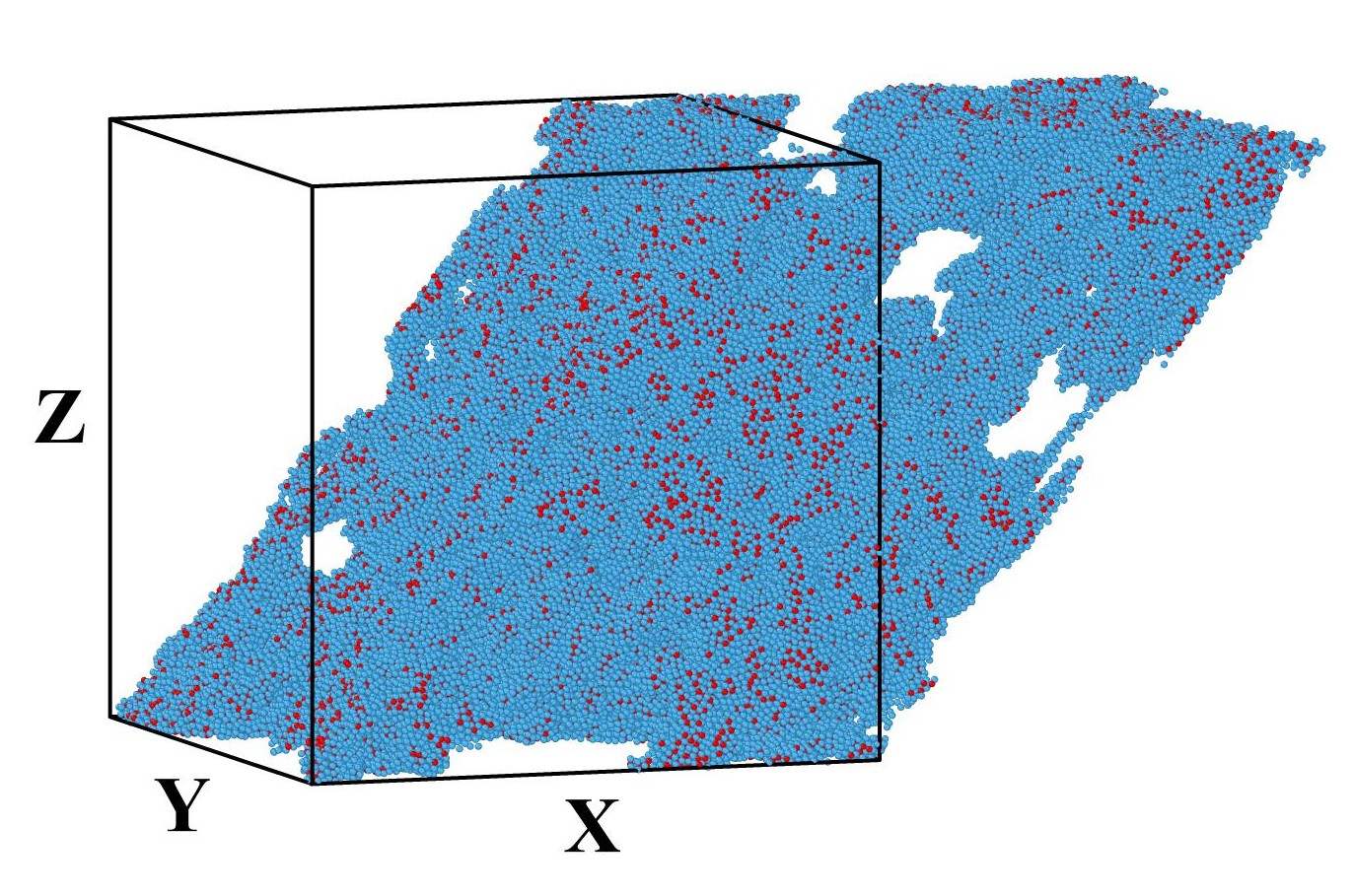}&
		\includegraphics[width=45mm]{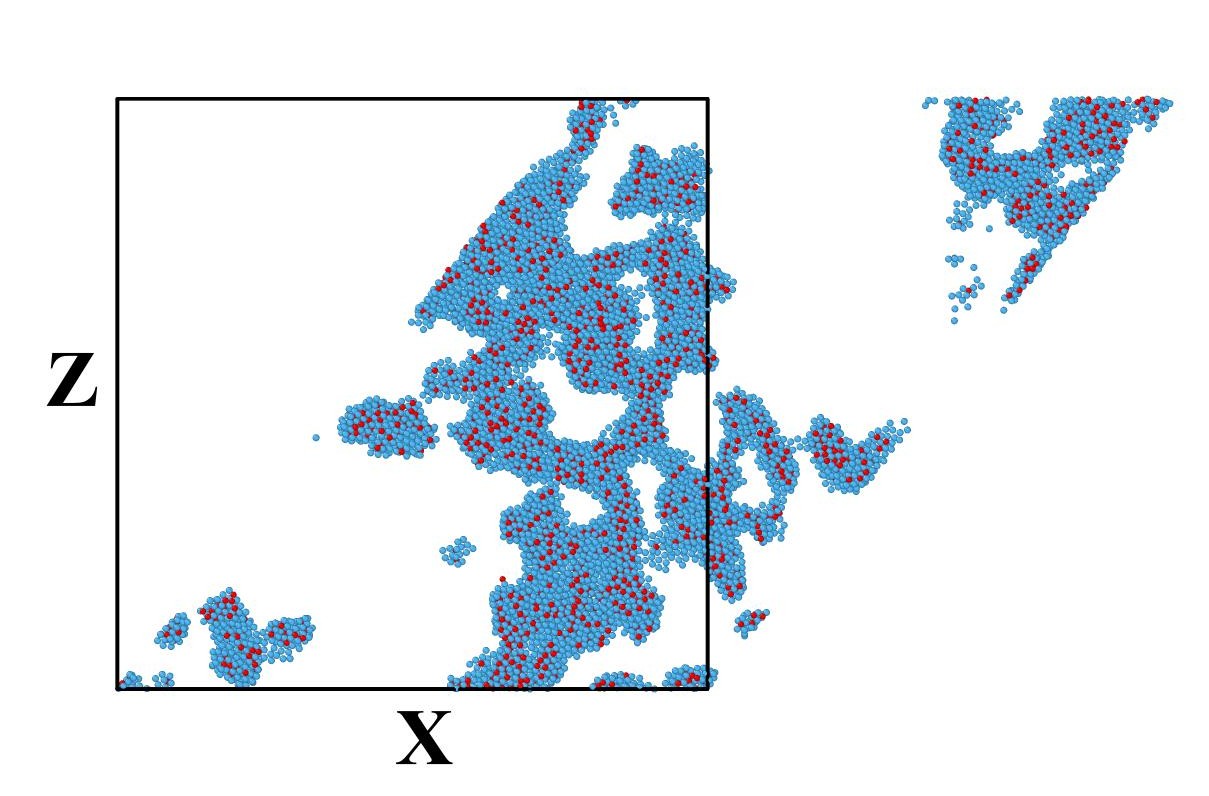}
	\end{tabular}
\caption{(a) Three-dimensional snapshots for the porous glasses for the average density $\rho=0.5$ and shear strain $\gamma = 0.2, 0.4, 0.6$ and $0.8$ (from top to bottom). (b) Two-dimensional cross sections of the snapshots in (a).}
\label{sheared-config-0.5}
\end{figure}

\begin{figure}[htp]
	\renewcommand{\thefigure}{8}
	\centering
	\begin{tabular}{cc}
		\centering
		\includegraphics[width=44mm]{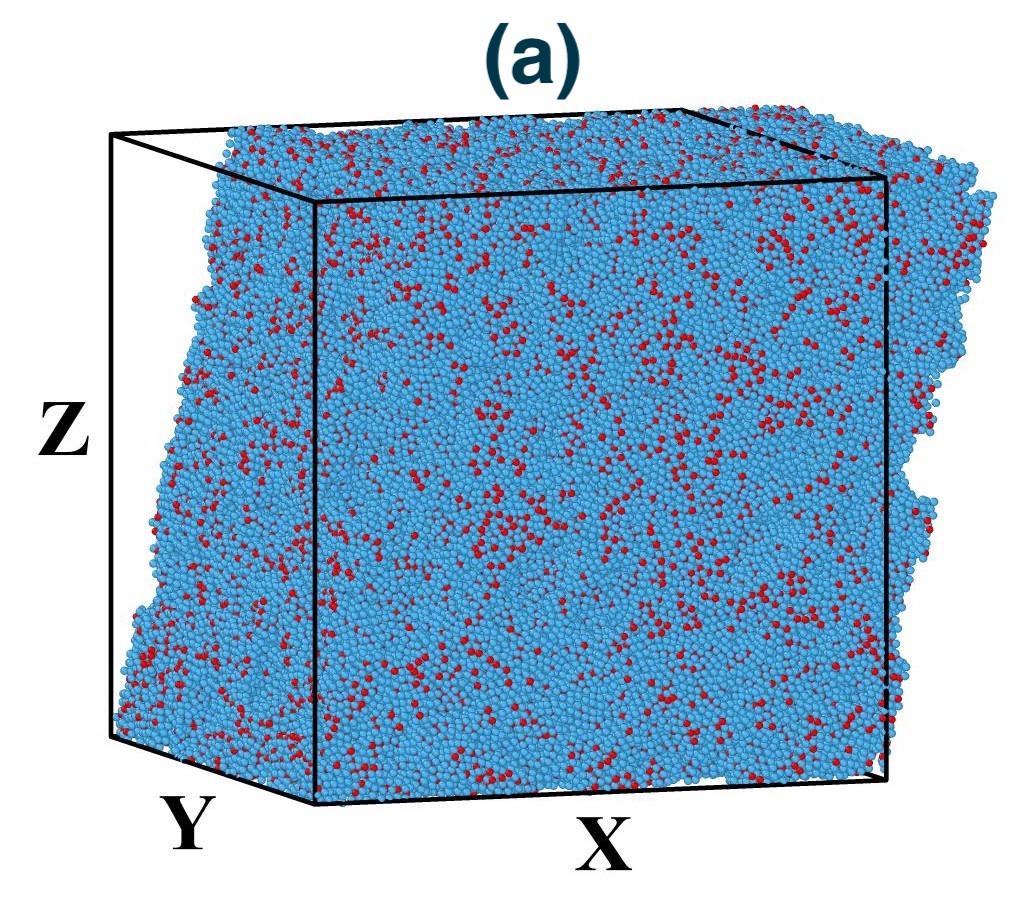}&
		\includegraphics[width=40mm]{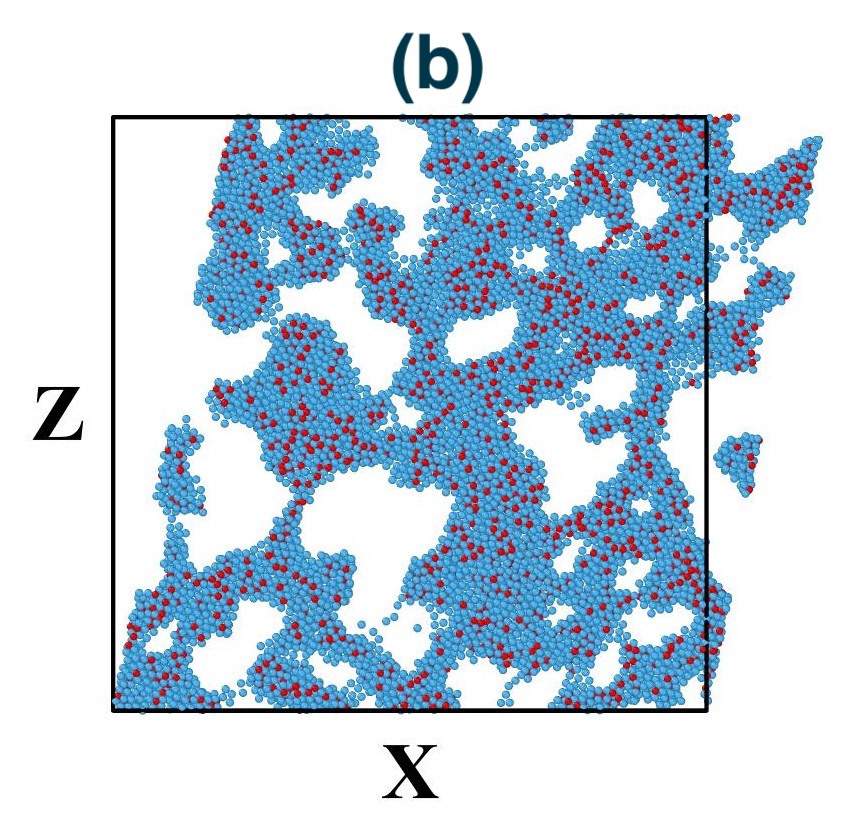}\\
		\includegraphics[width=40mm]{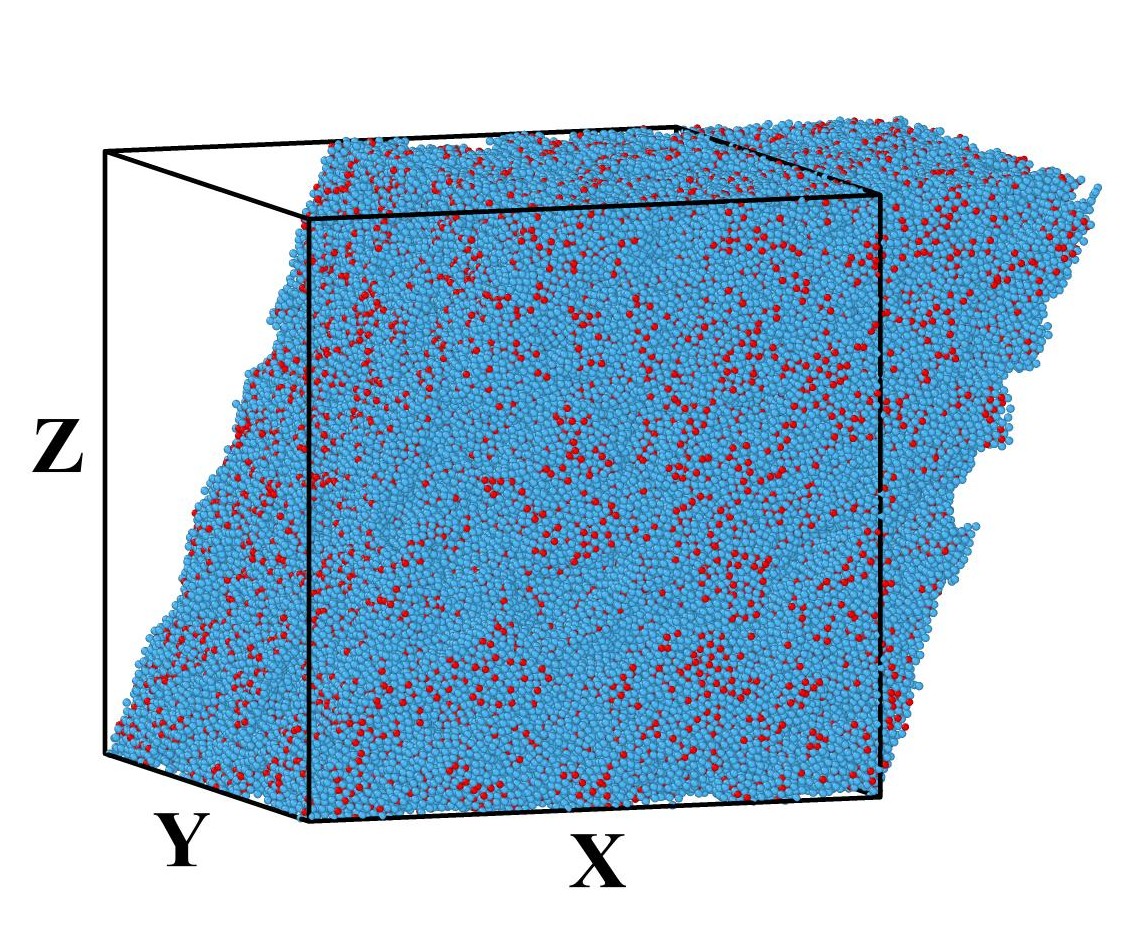}&
		\includegraphics[width=40mm]{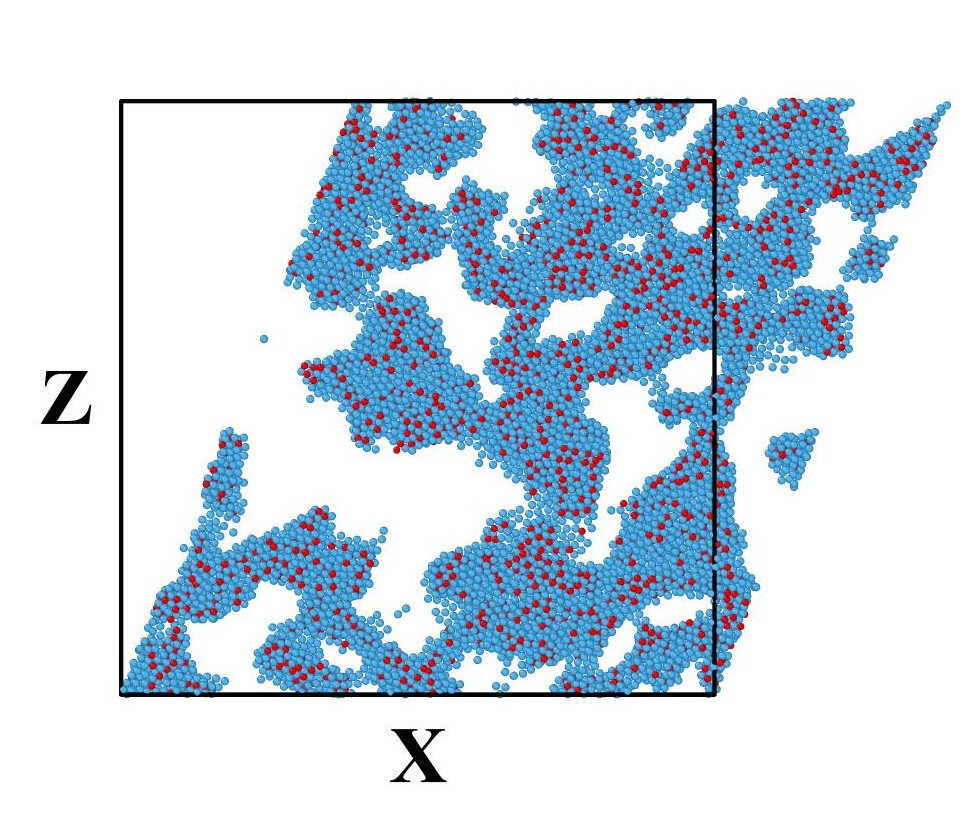}\\
		\includegraphics[width=40mm]{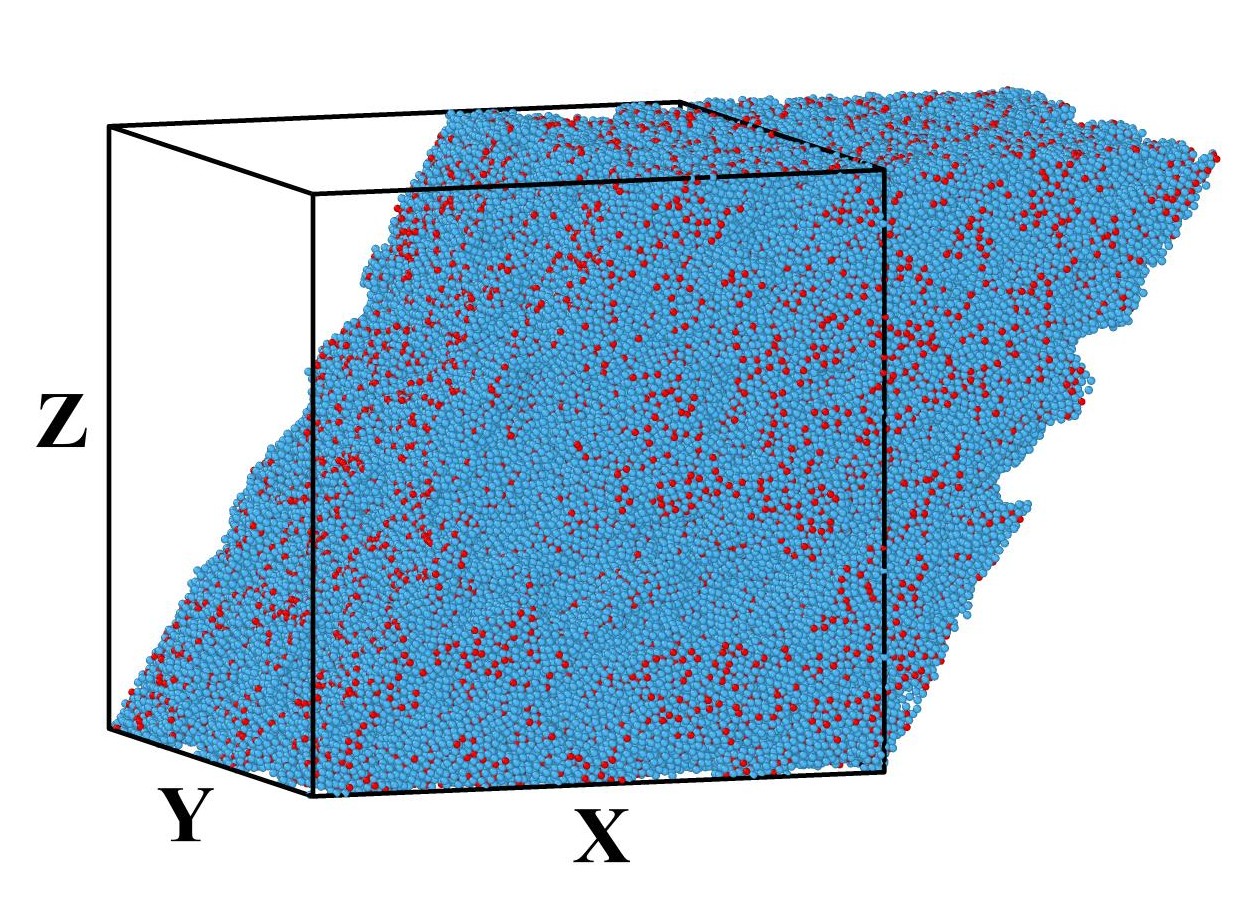}&
		\includegraphics[width=40mm]{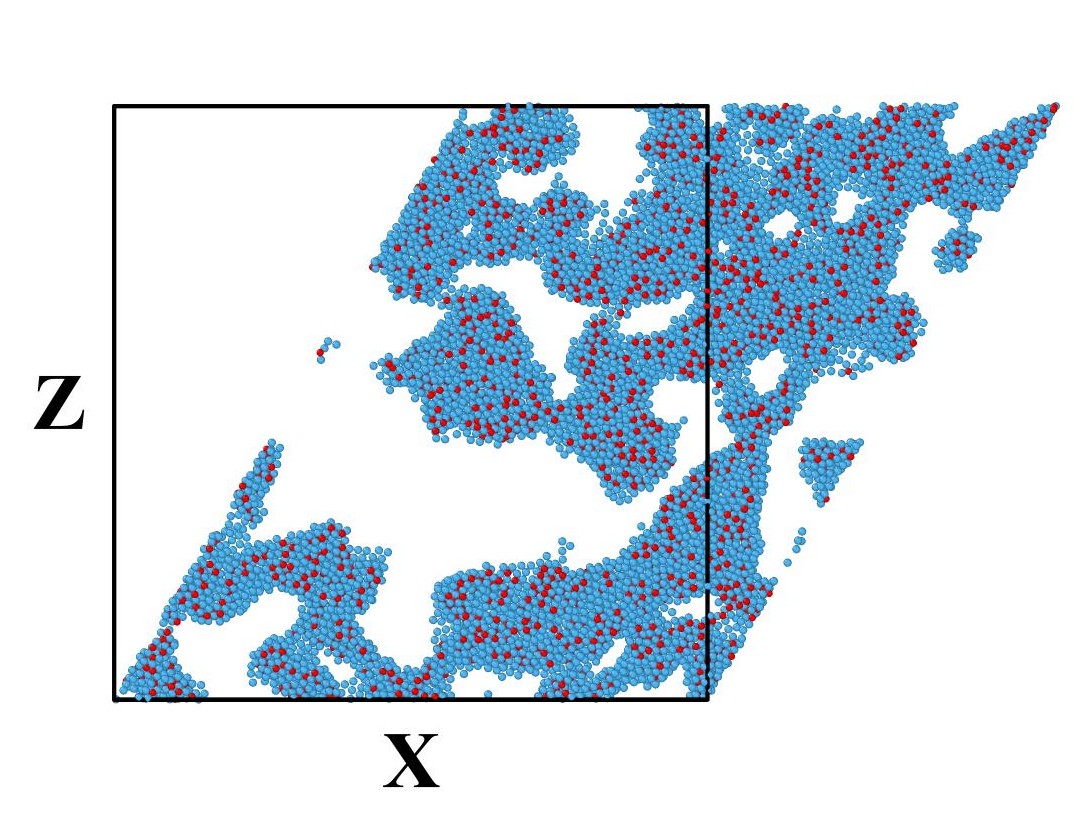}\\
		\includegraphics[width=40mm]{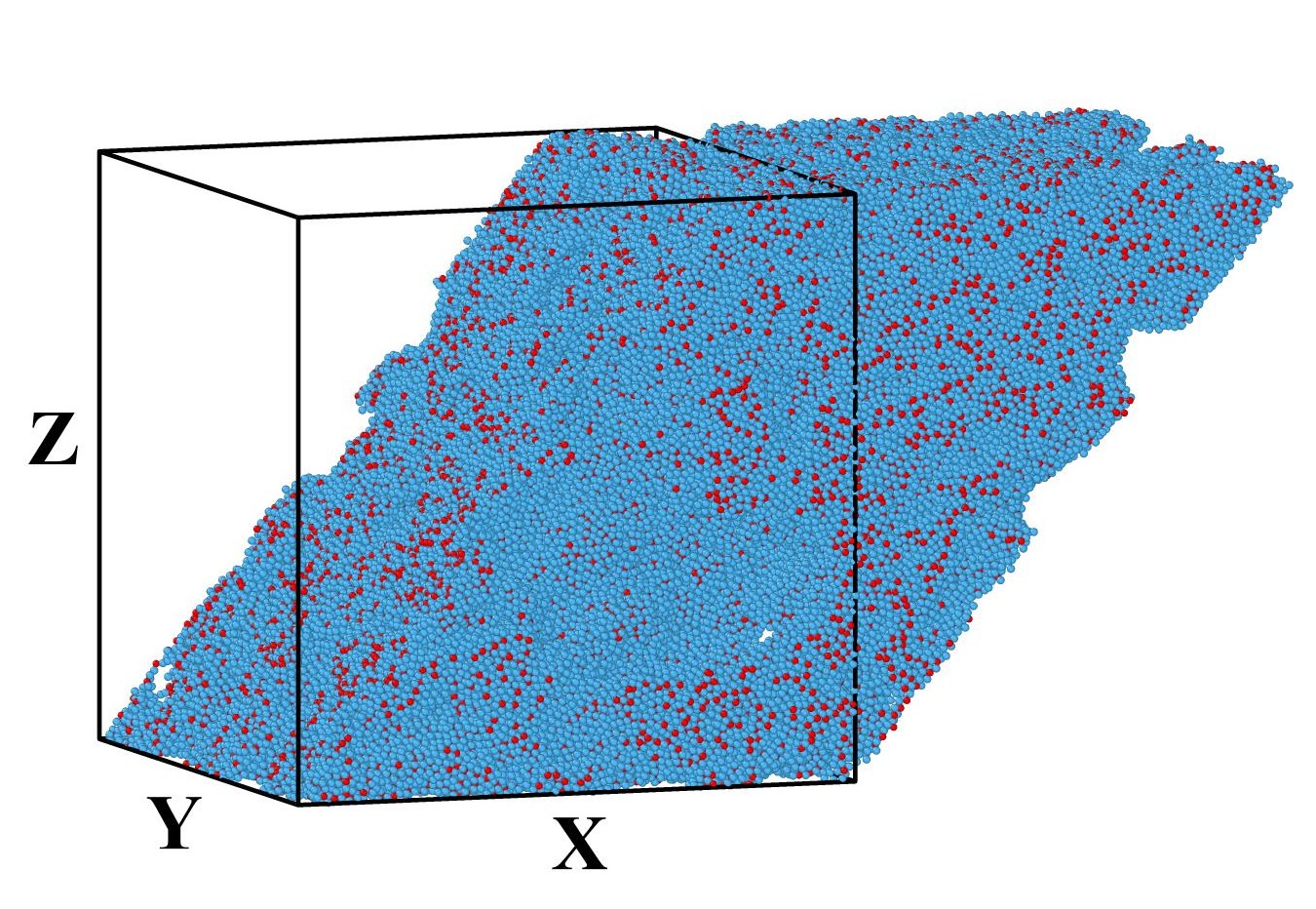}&
		\includegraphics[width=40mm]{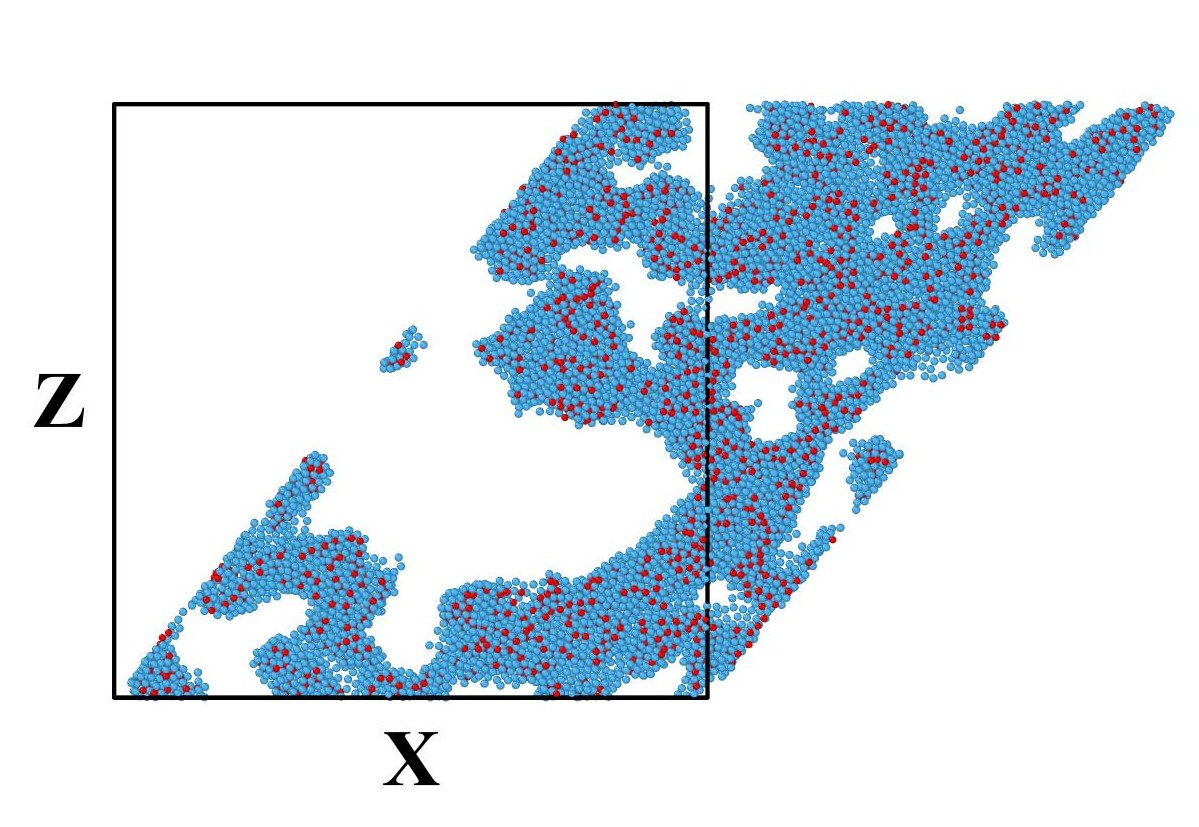}\\
	\end{tabular}
\caption{(a) Three-dimensional snapshots for the porous glasses for the average density $\rho=0.8$ and shear strain $\gamma = 0.2, 0.4, 0.6$ and $0.8$ (from top to bottom). (b) Two-dimensional cross sections of the snapshots in (a).}
	\label{sheared-config-0.8}
\end{figure}

The representative snapshots for the deformed porous glass samples obtained from our simulation are shown in Figs. \ref{sheared-config-0.2}a-\ref{sheared-config-0.8}a. for three different chosen densities $\rho=0.2, 0.5$ and $0.8$. For better visualization we also show the two-dimensional cross-section of the same configurations in Figs. \ref{sheared-config-0.2}b-\ref{sheared-config-0.8}b. 
We observe that the pore structure changes with shear. The effect of strain and the underlying mechanism for the structural change of pores can be understood in the following way. Due to sudden thermal quench at very low temperature the glassy sample under consideration exists in metastable states in the potential energy landscape. Due to the lack of thermal energy, the system can not overcome the potential energy barrier and can be envisaged as effectively confined in that minimum. As the shear is applied, the energy landscape deforms continuously and the system configuration follows the location of a single energy minimum. This process continues until the minimum where the system resides flattens out completely and hits the saddle point \cite{Lemaitre2006_PRE}. As a result, the system rolls down to the neighboring minimum which is accompanied by irreversible rearrangement of particles and pore size and structural change. This phenomenon is known as plastic instability. It is worth mentioning here that the linear regime of the stress-strain curve is punctuated with small plastic drops involving localized rearrangement of particles. As a result we do not see any appreciable change in pore structure in this regime. On further increase of strain, the material yields via the system spanning shear band formation with a large displacement of particles which induces significant change in the pore structure. Further insight on the deformed pore structure can be obtained by analyzing the distribution of the pore size which is discussed in the next section. 
\subsection{Pore size distribution and the scaling law}
\begin{figure*}[htp]
	\renewcommand{\thefigure}{9}
	\centering
	\begin{tabular}{cc}
		\centering
		\includegraphics[width=58mm]{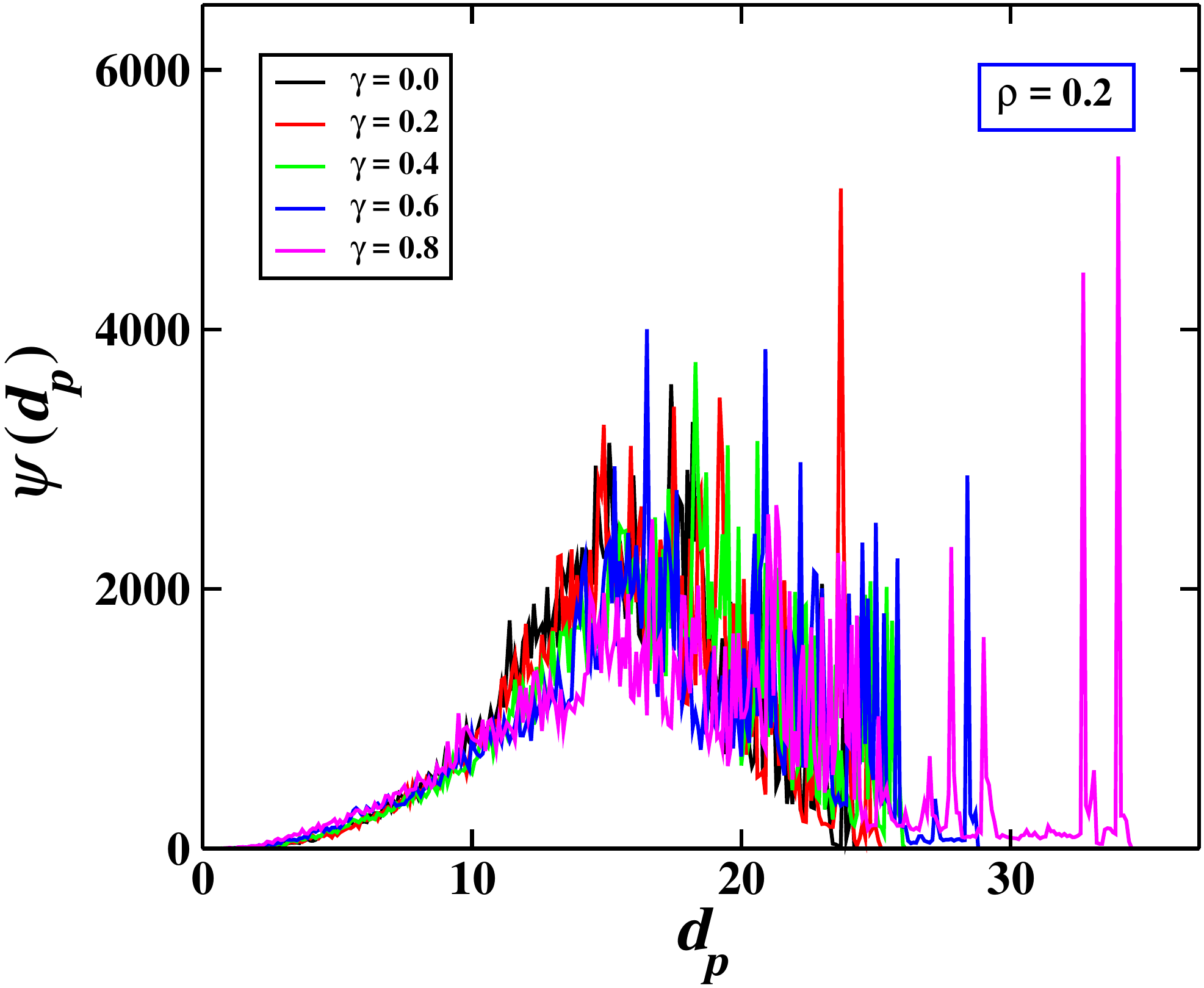}&
		\includegraphics[width=60mm]{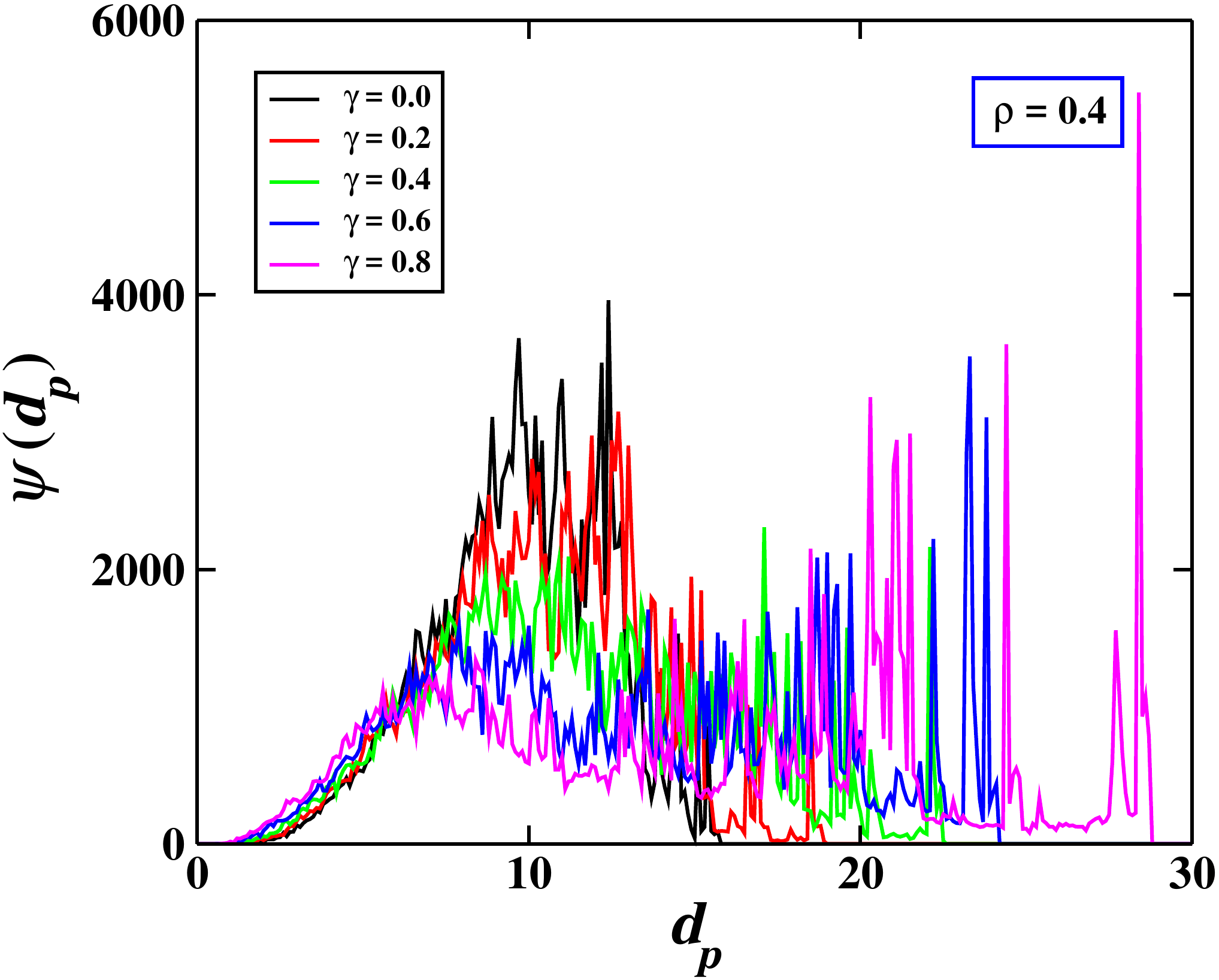}\\
		\includegraphics[width=60mm]{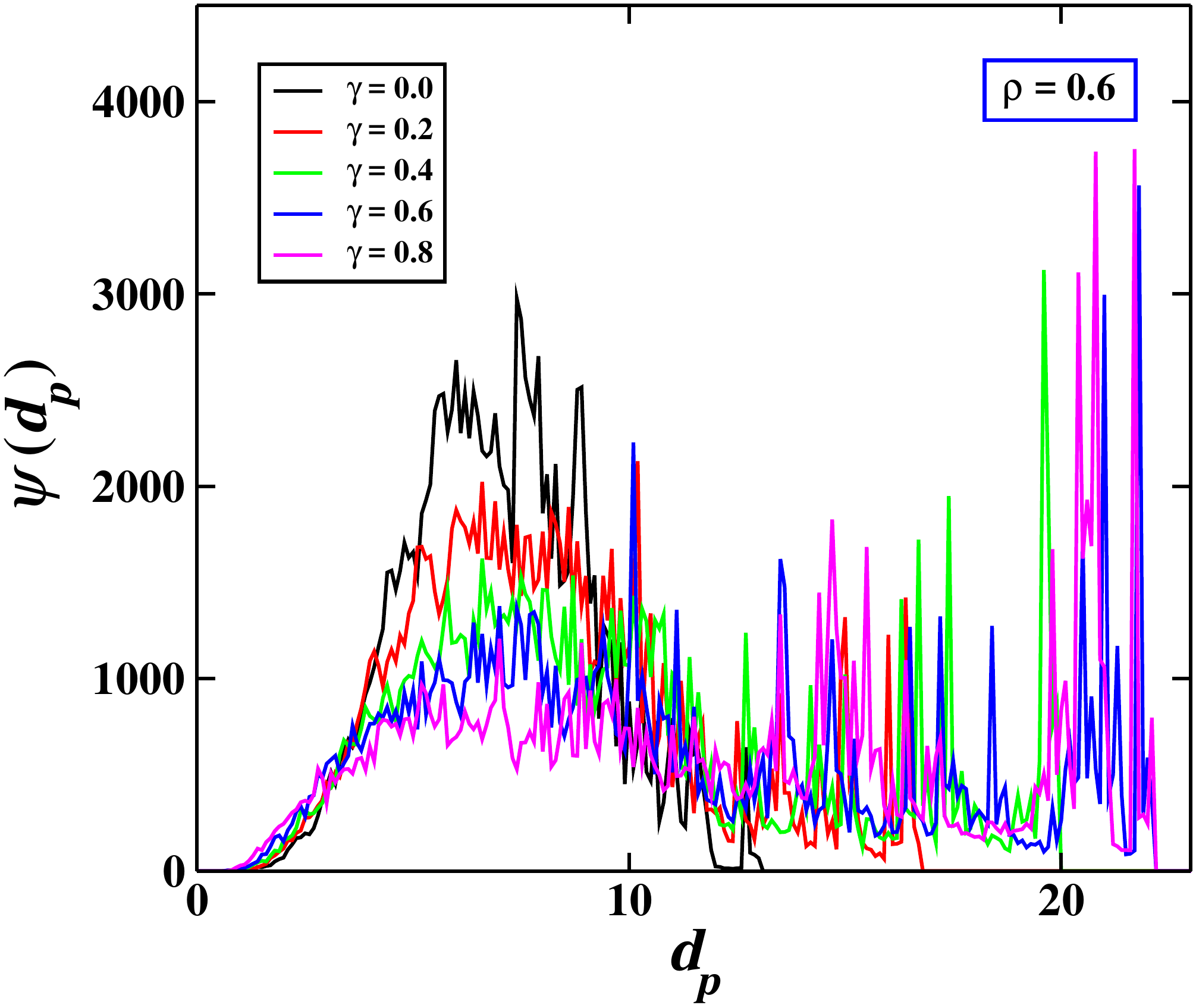}&
		\includegraphics[width=60mm]{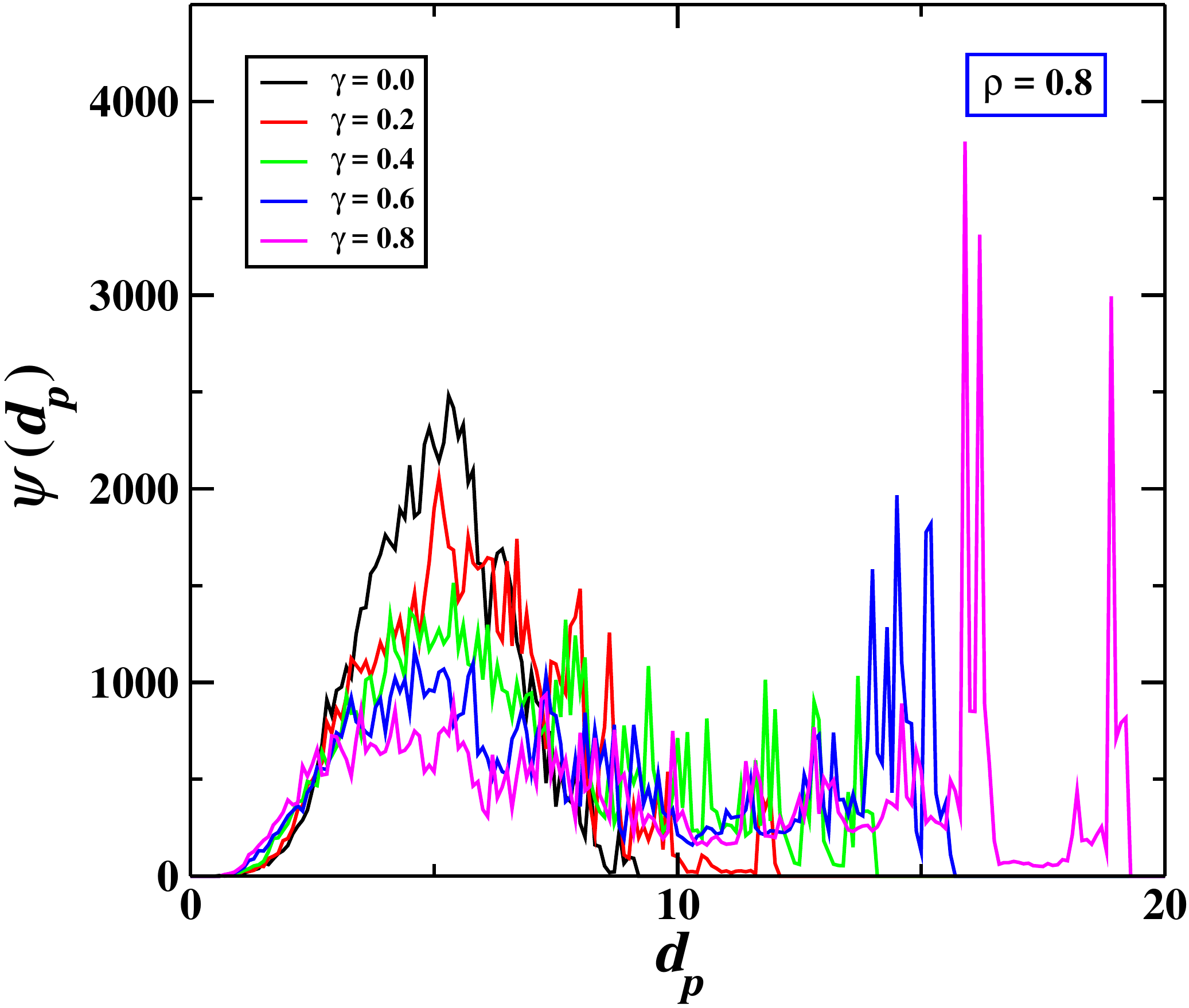}\\
	\end{tabular}
\caption{Typical pore size distribution function $\psi(d_p)$ for an individual sample for the densities $\rho=0.2, 0.4, 0.6$ and $0.8$ at various shear strain $\gamma$.}
	\label{psd-ind}
\end{figure*}

\begin{figure}[h!]
	\renewcommand{\thefigure}{10}
	\centering
	\includegraphics[width=0.99\columnwidth]{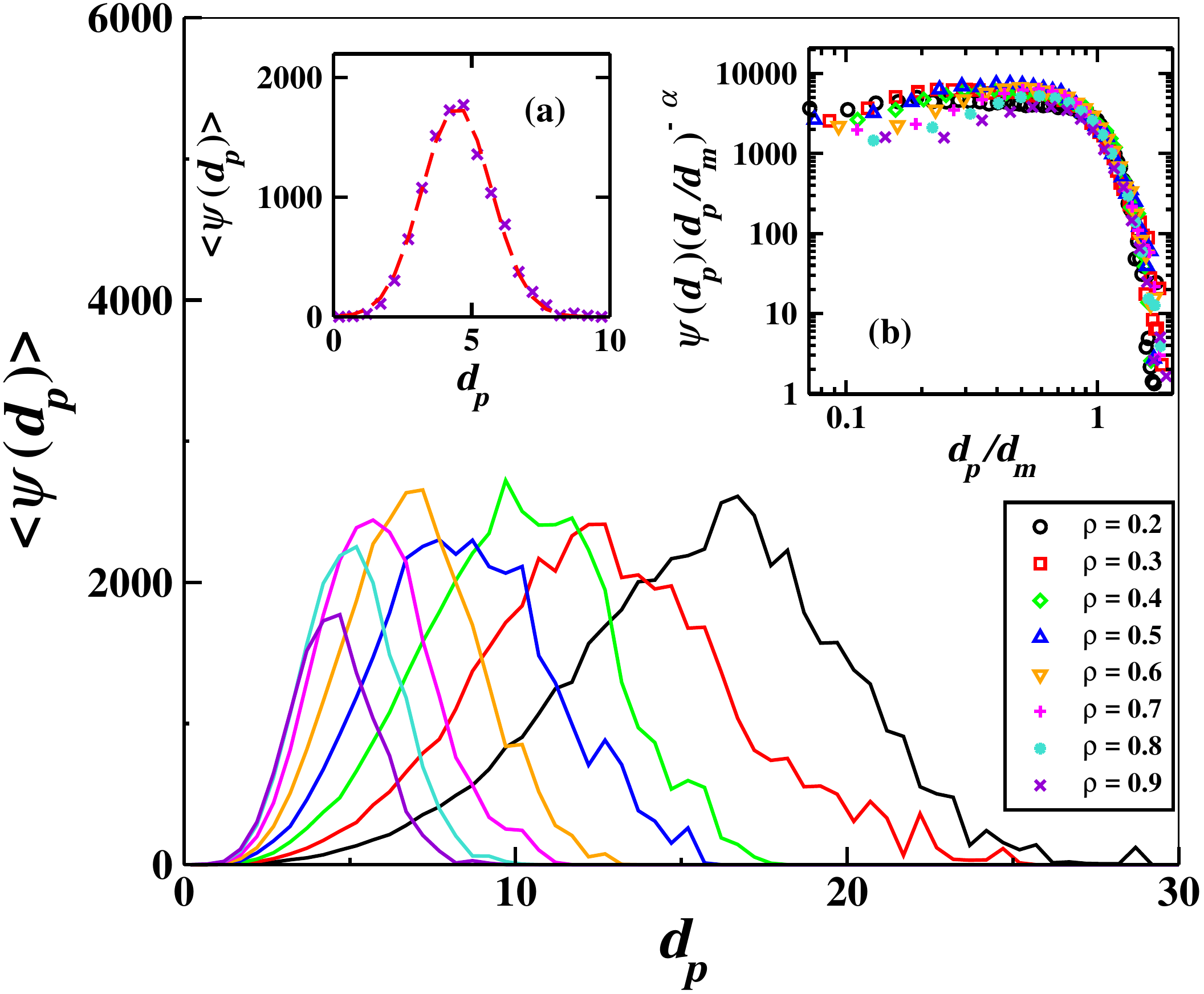}
	\caption{The averaged pore size distribution function $\langle\psi(d_p)\rangle$ for the undeformed system at densities $\rho=0.2, 0.3, 0.4, 0.5, 0.6, 0.7, 0.8$ and $0.9$. Inset (a): The $\langle\psi(d_p)\rangle$ data for $\rho=0.9$ fitted with Gaussian function. Inset (b): Data collapse of the $\langle\psi(d_p)\rangle$ data for all densities following Eq. 3. (see text)}
	\label{psd-undeformed}
\end{figure}

\begin{figure*}[htp]
	\renewcommand{\thefigure}{11}
	\centering
	\begin{tabular}{cc}
		\centering
		\includegraphics[width=60mm]{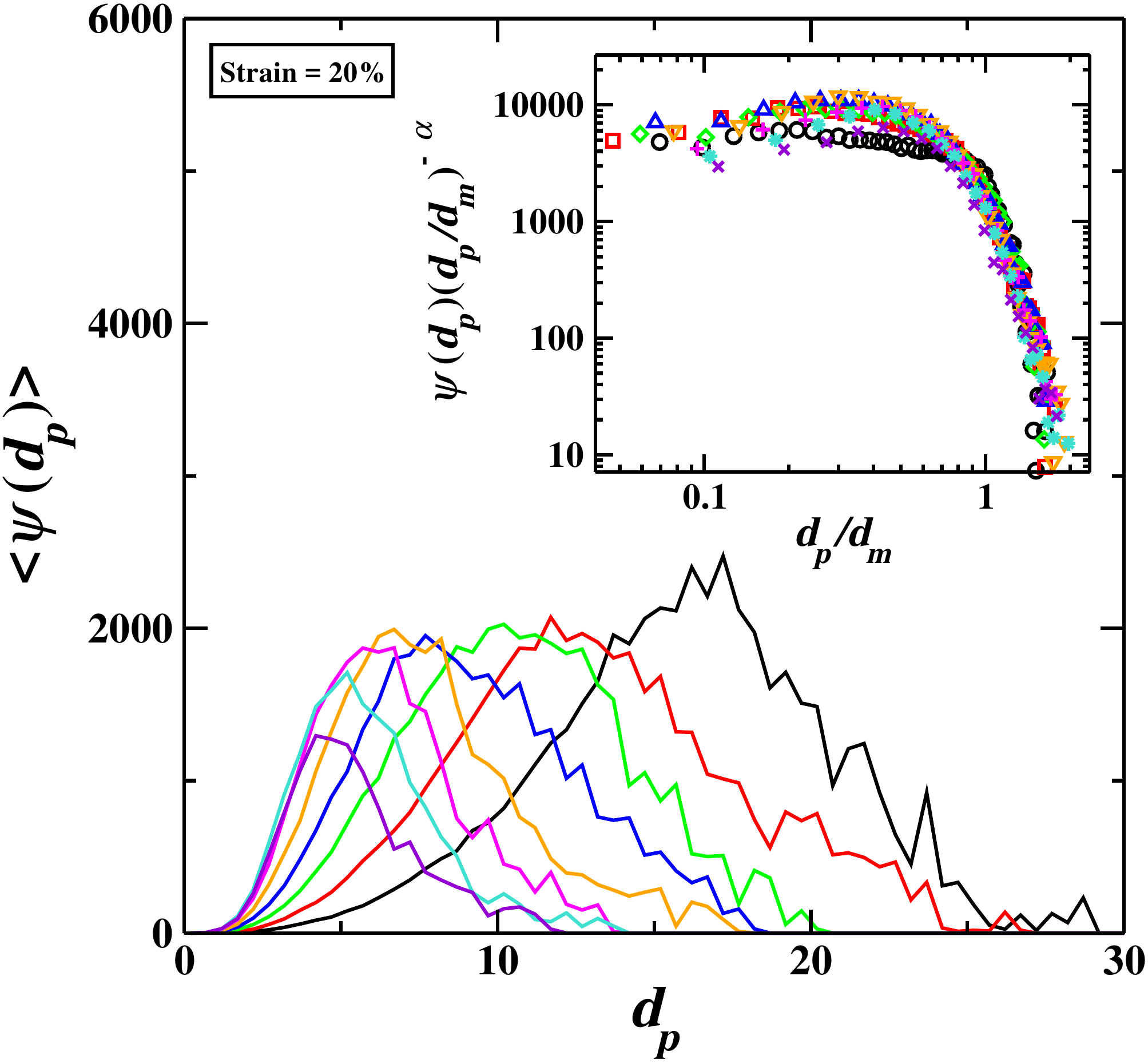}&
		\includegraphics[width=60mm]{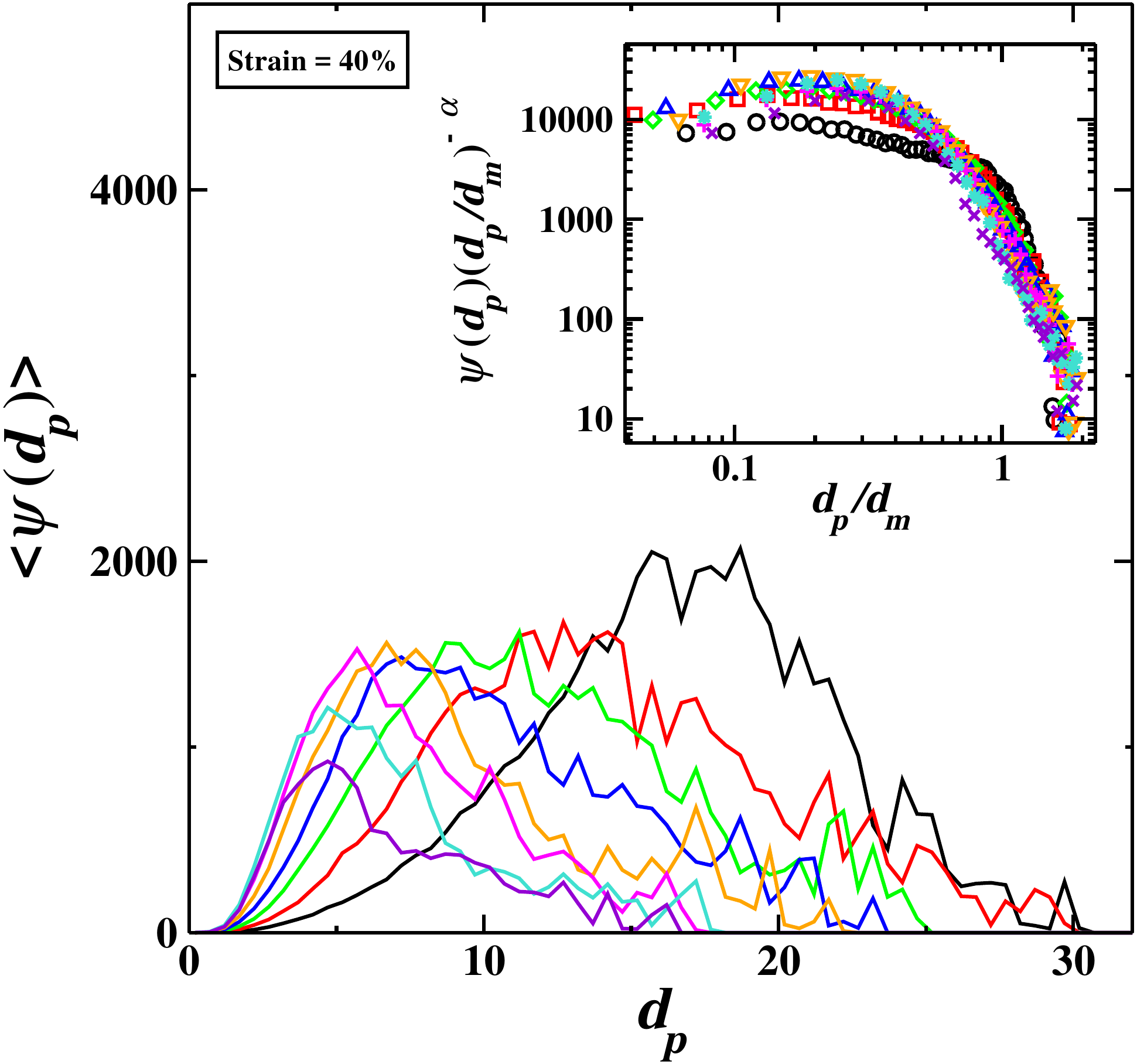}\\
		\includegraphics[width=60mm]{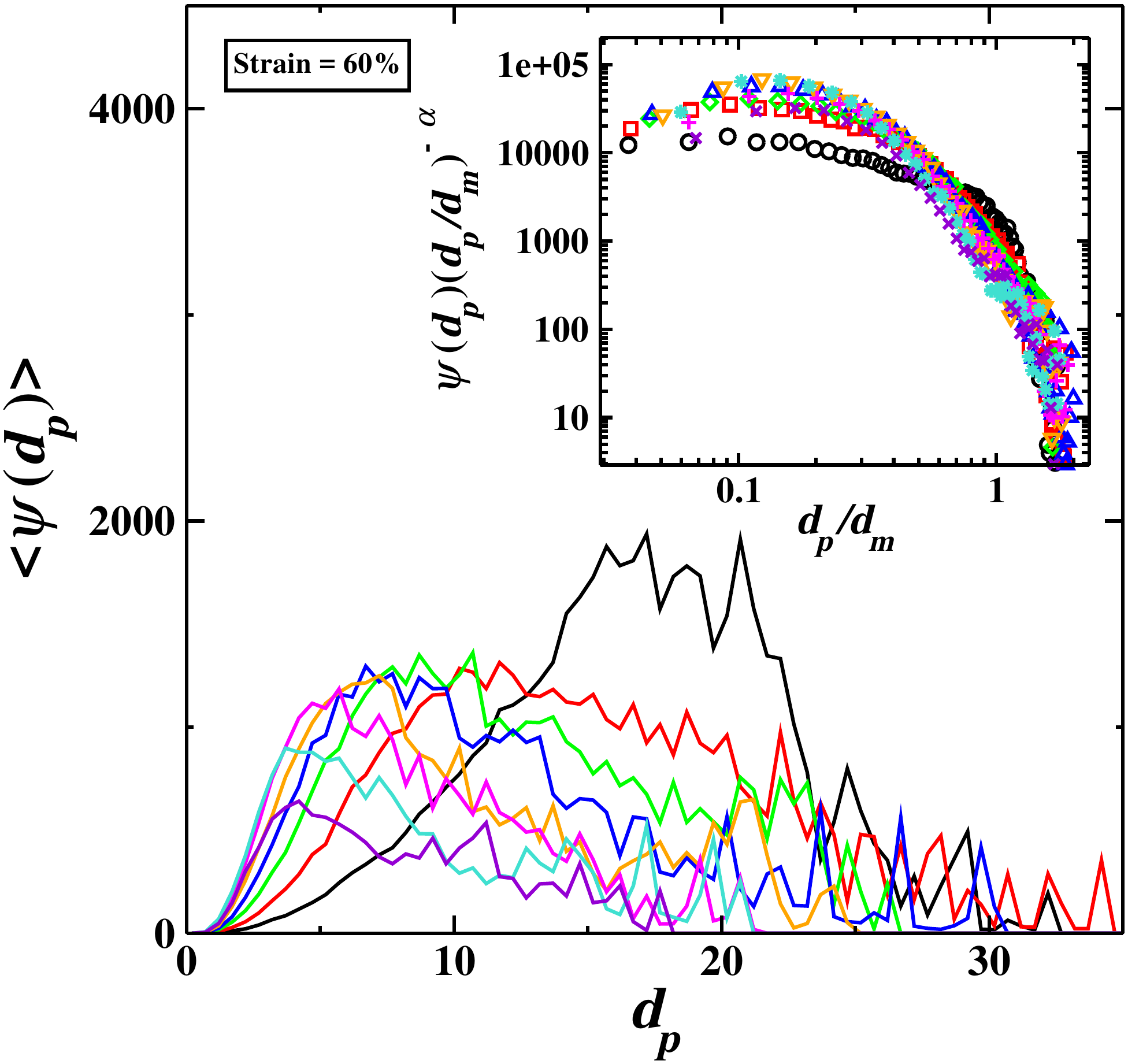}&
		\includegraphics[width=60mm]{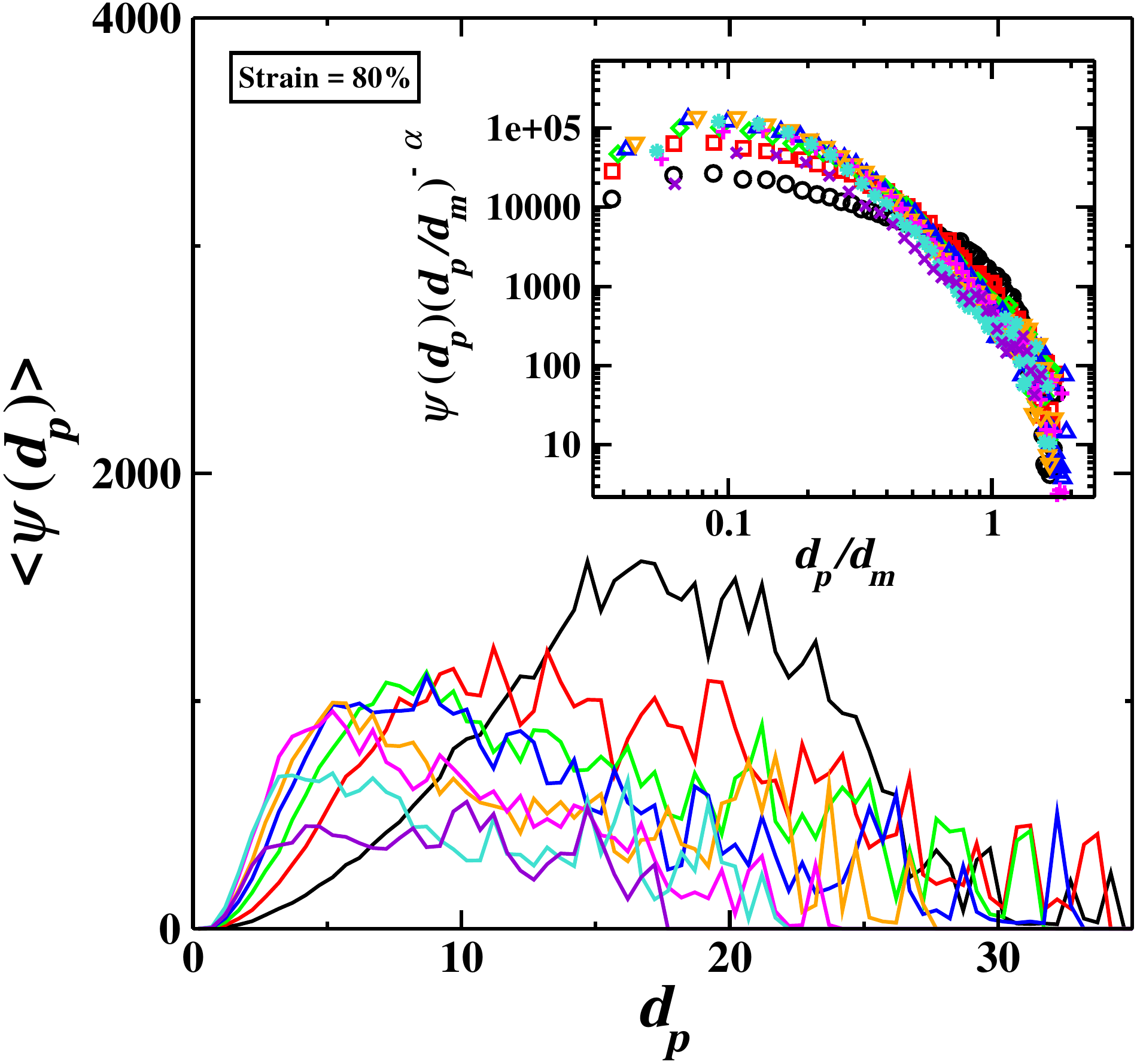}\\
	\end{tabular}
\caption{The averaged pore size distribution function $\langle\psi(d_p)\rangle$ for the deformed system at strain $\gamma=0.2, 0.4, 0.6,$ and $0.8$ for the densities $\rho=0.2, 0.3, 0.4, 0.5, 0.6, 0.7, 0.8$ and $0.9$.  Inset (b): Data collapse of the $\langle\psi(d_p)\rangle$ data for all densities following Eq. 3. (see text). The same color code as in Fig. \ref{psd-undeformed}}
	\label{psd-sheared}
\end{figure*}

The effect of shear on the pore structure is best realized in terms of pore size distribution function denoted as $\psi(d_p)$ where $d_p$ is the pore size. The $\psi(d_p)$ is computed using the open-source Zeo++ software \cite{zeo1, zeo2, zeo3}. This is displayed in Fig. \ref{psd-ind} where we show the sheared configurations for different densities in the range $0.2 \le \rho \le0.9$. 
From the results, it is evident that the $\psi(d_p)$ broadens with decrease in density. Under the effect of shear, the distribution becomes skewed towards the higher value of the pore size. This clearly indicates that as the system deforms, some of the pores start merging and larger pores are developed. This is reflected in Fig. \ref{psd-ind} by the large peaks towards the higher $d_p$ at high deformation. Therefore, the effect of shear can be comprehended as the coalition of small pores and the formation of some large dominant pores and the material solidification.\\
Figs. \ref{psd-undeformed} and \ref{psd-sheared} summarize the effect of mechanical strain on the pore size distribution $\langle\psi(d_p)\rangle$ computed at different densities and strain values. The angular brackets represent the ensemble averaging over $5$ samples. We observe the distribution curves become wider with decrease in density and the peak shifts towards the higher $d_p$ value. Careful examination reveals that in the undeformed case ($\gamma=0$) at high density $\rho=0.9$ and $1.0$ (not shown here), the distribution curve is symmetric around the peak and exhibits a Gaussian nature. This is demonstrated in Fig. \ref{psd-undeformed} by fitting the data with Gaussian distribution. 
Our result is consistent with the recent experimental observation \cite{yavari1}. This behavior changes completely for $\rho \le 0.8$ where the peak gradually shifts towards the higher pore sizes with large interconnected pores extended up to the system size. Therefore, based on the characteristic behavior of the distribution function, the system under consideration can be broadly categorized as bulk- and porous-type, and $\rho=0.9$ can be recognized as the transition density. 

Next we investigate if there exists any underlying universal behavior for the pore size distribution functions at different densities. For that we consider the following scaling ansatz \cite{maxim,nakayama}:
\begin{equation}
\label{ansatz}
\psi(d_p/d_m) = \left(d_p/d_m\right)^{-\alpha} \Phi(d_p). 
\end{equation}
Here $d_m$ represents the mean pore diameter computed as 
\begin{equation}
d_m= \sum_{s} n_s(d_p) d_{p(s)}^2/\sum_{s} n_s(d_p) d_{p(s)}
\end{equation}
where $n_s(d_p)$ is the number of pores with size $d_p$. $\alpha$ is the scaling exponent which is used as the fitting parameter. The quality of scaling obtained using Eq. \ref{ansatz} for all the densities is depicted in the inset of Fig. \ref{psd-undeformed}. A convincing data collapse is observed for the densities $\rho\le0.8$. The scaling process yields the fitting exponent as $\alpha=3$. The same scaling exponent was obtained when the porous glass was deformed at nonzero temperature \cite{maxim}. Therefore, $\alpha=3$ shows universal behavior, oblivious to thermal fluctuation.  At high density ($\rho=0.9$), the distribution curve deviates from the scaling law at the intermediate and smaller pore size region. This deviation confirms our previous understanding that the bulk type material with low porosity behaves differently from the highly porous material. The $\rho=0.9$, therefore, can be considered as the critical density of transition from porous to bulk type behavior.

The distribution $\langle\psi(d_p)\rangle$ for the deformed states are shown in Fig. \ref{psd-sheared} for all density values.
As mentioned above, the deformation significantly modifies the distribution with small pores merging and forming larger pores. Therefore, the $\langle\psi(d_p)\rangle$ becomes flattered towards the larger pore size. As a result, the distribution functions start to deviate from the scaling law with increasing strain value. This is shown in the inset of the respective figures. Proper scaling of the pore size distribution at finite deformation requires further investigation with the strain-dependent function $\Phi(d_p/d_m,\gamma)$. This is beyond the scope of this paper and will be reported elsewhere.  
\section{Summary and Conclusions}
In summary, we have studied a modeled porous glass prepared by rapid thermal quench from high temperature using atomistic computer simulation. A wide range of pore structures was created by varying the average density of the glassy system. Visual inspection revealed interconnected porous networks at low densities whereas randomly distributed isolated pores at high densities. Furthermore, the evolution of the pore structure was investigated under simple shear deformation in the athermal quasistatic limit which is free from any thermal noise. We observed the shear modulus strongly depends on density. The porosity dependence of shear modulus computed from our model follows the prediction of percolation theory. Under mechanical loading, the topography of the pore structure changed significantly with strain. This was illustrated by computing the pore size distribution function.  We found that there exists a critical density above which the material behaves as bulk-type with the pore size distribution showing Gaussian nature. A scaling law was offered for the distribution function below this critical density where the material behaves as porous-type. The scaling exponent turned out to be universal and independent of temperature. A major effect of shear strain on the pore structure is the broadening of distribution curve by coalition of pores. As a result, the scaling law is found to deviate with increase in deformation which requires further investigation. We believe that the present results should have further implications for the study and understanding of the structural and mechanical properties of porous glasses. At this point the full understanding of the scaling behavior of the pore size distribution function in the deformed states is not available and it will certainly be worthwhile to find it.

\section*{Acknowledgment}
We acknowledge Science and Engineering Research Board (SERB), Department of Science and Technology (DST), Government of India (no. SRG/2019/001923) for financial support.


\section*{References} 
\begin{thebibliography}{}
\bibitem{yang}	
X.-Y. Yang, L.-H. Chen, Y. Li, J. C. Rooke, C. Sanchez, and B.-L. Su, Chem. Soc. Rev. \textbf{46}, 481 (2017).
\bibitem{porter}
M.M. Porter, R. Imperio, M. Wen, M.A. Meyers, J. McKittrick, Adv. Funct. Mater. \textbf{24}, 1978 (2014).
\bibitem{Li}
Y. Li, Z.-Y. Fu, B.-L. Su, Adv. Funct. Mater. \textbf{22}, 4634 (2012).
\bibitem{zhang}
J. Zhang, H. Zhou, K. Yang, Y. Yuan, C. Liu, Biomaterials \textbf{34}, 9381 (2013).
\bibitem{hammel}
E.C. Hammel, O.L.-R. Ighodaro, O.I. Okoli, Ceram. Int. \textbf{40}, 15351 (2014).
\bibitem{fu}
Q. Fu, E. Saiz, A.P. Tomsia, Adv. Funct. Mater. \textbf{21}, 1058 (2011).
\bibitem{sarac}
B. Sarac, B. Klusemann, T. Xiao, S. Bargmann, Acta Mater. \textbf{77}, 411 (2014).
\bibitem{sarac1}
B. Sarac, D. Sopu, E. Park, J.K. Hufenbach, S. Oswald, M. Stoica, J. Eckert, Metals \textbf{5}, 920
(2015).
\bibitem{sopu}
D. Sopu, C. Soyarslan, B. Sarac, S. Bargmann, M. Stoica, J. Eckert, Acta Mater. \textbf{106}, 199 (2016).
\bibitem{wang1}
J. Wang, P. D. Hodgson, J. Zhang, W. Yan, and C. Yang, Comput. Mater. Sci. \textbf{50}, 211 (2010).
\bibitem{campbell}
T. Campbell, R. K. Kalia, A. Nakano, F. Shimojo, K. Tsuruta, P. Vashishta, and S. Ogata, Phys. Rev. Lett. \textbf{82}, 4018 (1999).
\bibitem{rimsza}
J. M. Rimsza and J. Du, J. Am. Ceram. Soc. \textbf{97}, 772 (2014).
\bibitem{chen}
Y.-C. Chen, Z. Lu, K. Nomura, W. Wang, R.K. Kalia, A. Nakano, P. Vashishta, Phys. Rev. Lett. \textbf{99}, 155506 (2007).
\bibitem{testard}
V. Testard, L. Berthier, and W. Kob, Phys. Rev. Lett. \textbf{106}, 125702 (2011).
\bibitem{testard1}
V. Testard, L. Berthier, and W. Kob, J. Chem. Phys. \textbf{140}, 164502 (2014).
\bibitem{yavari}
A. R. yavari, A. Le Moulec, A. Inoue, N. Nishiyama, N. Lupu,
E. Matsubara, W. J. Botta, G. Vaughan, M. Di Michiel, and A. Kvick, Acta Mater. \textbf{53}, 1611 (2005).
\bibitem{maxim}
Maxim A. Makeev and Nikolai V. Priezjev Phys. Rev. E \textbf{97}, 023002 (2018).
\bibitem{ongari}
D. Ongari, P.G. Boyd, S. Barthel, M. Witman, M. Haranczyk, B. Smit, Langmuir
\textbf{33}, 14529 (2017).
\bibitem{michele}
L. Di Michele, D. Fiocco, F. Varrato, S. Sastry, E. Eisera, and G. Foffi, Soft Matter \textbf{10}, 3633 (2014).
\bibitem{kob}
W. Kob, H.C. Andersen, Phys. Rev. E \textbf{51}, 4626 (1995).
\bibitem{verlet}
L. Verlet, Phys. Rev. \textbf{159}, 98 (1967).
\bibitem{nose}
D. Frenkel and B. Smit, Academic Press, (2002). 
\bibitem{gaurav}
G.P. Shrivastav, P. Chaudhuri, J. Horbach, Phys. Rev. E \textbf{94}, 042605 (2016).
\bibitem{gaurav1}
G.P. Shrivastav, P. Chaudhuri, J. Horbach, Journal of Rheology \textbf{60}, 835 (2016).
\bibitem{Liu}
C. Liu, E.E. Ferrero, F. Puosi, J.-L. Barrat, K. Martens, Phys. Rev. Lett. \textbf{116}, 065501 (2016).
\bibitem{Lacks1999_JCP}
D.L. Malandro, D.J. Lacks, Journal of Chemical Physics \textbf{110}, 4593 (1999).
\bibitem{Barrat2002_PRB}
A. Tanguy, J.P. Wittmer, F. Leonforte, J.-L. Barrat, Phys. Rev. B \textbf{66}, 174205 (2002).
\bibitem{Lemaitre2004_PRL1}
C. Maloney, A. Lema\^{i}tre, Phys. Rev. Lett. \textbf{93}, 016001 (2004).
\bibitem{Lemaitre2004_PRL2}
C. Maloney, A. Lema\^{i}tre, Phys. Rev. Lett. \textbf{93}, 195501 (2004).
\bibitem{Lemaitre2006_PRE}
C. Maloney, A. Lema\^{i}tre, Phys. Rev. E \textbf{74}, 016118 (2006).
\bibitem{Procaccia2009_PRE}
E. Lerner, I. Procaccia, Phys. Rev. E \textbf{79}, 066109 (2009). 
\bibitem{Procaccia2010_PRE}
S. Karmakar, E. Lerner, I. Procaccia, Phys. Rev. E \textbf{82}, 055103R (2010). 
\bibitem{Procaccia2011_PRE}
H.G.E. Hentschel, S. Karmakar, E. Lerner, I. Procaccia, Phys. Rev. E \textbf{83}, 061101 (2011).
\bibitem{santhosh}
S. Kumar and B. S. Gupta, Journal of Physics: Cond. Mat. Accepted Manuscript (2021)
\bibitem{lammps}
S. J. Plimpton, J. Comp. Phys. \textbf{117}, 1 (1995).
\bibitem{yavari1}
A. R. Yavari, A. Le Moulec, A. Inoue, N. Nishiyama, N. Lupu, E. Matsubara, W. J. Botta, G. Vaughan, M. Di Michiel, and A. Kvick, Acta Mater. \textbf{53}, 1611 (2005).
\bibitem{sahimi}
M. Sahimi, Taylor and Francis, (1994).
\bibitem{phani}
K. K. Phani and S. K. Niyogi, J. Mater. Sci. \textbf{22}, 257 (1987).
\bibitem{kov}
J. Kovacik, Journal Of Materials Science Letters \textbf{20}, 1953 (2001).
\bibitem{kov1}
J. Kovacik and F. Simancik, Scripta Mater. \textbf{39}, 239 (1998).
\bibitem{zeo1}
R.L. Martin, B. Smit, and M. Haranczyk, J. Chem. Inf. Model. \textbf{52}, 308 (2012).
\bibitem{zeo2}
T.F. Willems, C.H. Rycroft, M. Kazi, J.C. Meza, and M. Haranczyk, Mesopor. Mater. \textbf{149} 134 (2012).
\bibitem{zeo3}
D. Ongari, P.G. Boyd, S. Barthel, M. Witman, M. Haranczyk, and B. Smit, Langmuir \textbf{33} 14592 (2017).
\bibitem{nakayama}
T. Nakayama, K. Yakubo, and R. L. Orbach, Rev. Mod. Phys. \textbf{66}, 381 (1994).

\end{thebibliography}
\end{document}